\documentclass[aps,prl,twocolumn,amsmath,amssymb,nofootinbib,superscriptaddress,floatfix,reprint,longbibliography]{revtex4-1}
\usepackage[dvips]{graphicx}
\usepackage{latexsym}
\usepackage{amsmath}
\usepackage{amsfonts}
\usepackage{amssymb}
\usepackage{bm}
\usepackage{color}
\usepackage{txfonts}
\usepackage{float}
\usepackage{url}
\usepackage[colorlinks=true, urlcolor=blue, linkcolor=blue, citecolor=blue, pdftex]{hyperref}
\usepackage{xcolor}
\usepackage{ulem}
\usepackage{physics}
\usepackage{hhline}
\newif\ifmarkedmanuscript
  \markedmanuscripttrue
  \ifmarkedmanuscript
    
    \newcommand{\revdel}[1]{{\color{red}\sout{#1}}}
    
  \else
    
    \newcommand{\revdel}[1]{}
    
  \fi

\begin{document}
	\newcommand{\fig}[2]{\includegraphics[width=#1]{#2}}
	\newcommand{\pprl}{Phys. Rev. Lett. \ }
	\newcommand{\pprb}{Phys. Rev. {B}}

\title {Anisotropic vortex motion and two-dimensional superconducting transition}

\author{Zhipeng Xu}
\affiliation{Beijing National Laboratory for Condensed Matter Physics and Institute of Physics,
	Chinese Academy of Sciences, Beijing 100190, China}
\affiliation{School of Physical Sciences, University of Chinese Academy of Sciences, Beijing 100049, China}

\author{Kun Jiang}
\email{jiangkun@iphy.ac.cn}
\affiliation{Beijing National Laboratory for Condensed Matter Physics and Institute of Physics,
	Chinese Academy of Sciences, Beijing 100190, China}
\affiliation{School of Physical Sciences, University of Chinese Academy of Sciences, Beijing 100049, China}

\author{Jiangping Hu}
\email{jphu@iphy.ac.cn}
\affiliation{Beijing National Laboratory for Condensed Matter Physics and Institute of Physics,
	Chinese Academy of Sciences, Beijing 100190, China}
\affiliation{School of Physical Sciences, University of Chinese Academy of Sciences, Beijing 100049, China}
 \affiliation{New Cornerstone Science Laboratory, Institute of Physics, Chinese Academy of Sciences, Beijing 100190, China}

\date{\today}

\begin{abstract}
Vortex motion plays a central role in determining the resistance of two-dimensional superconductors, both in the context of the Berezinskii-Kosterlitz-Thouless (BKT) transition and in the mixed state of type-II superconductors under magnetic fields. In this study, we introduce an anisotropic pinning potential to investigate vortex-induced resistance across the BKT transition and the upper critical field \( H_{c2} \) transition. Our results demonstrate that the anisotropic pinning potential gives rise to distinct critical temperatures and upper critical fields along two orthogonal directions of current transport. These findings provide a general route toward the realization of multiple ``critical temperatures" in two-dimensional superconductors.
\end{abstract}
%\pacs{}
\maketitle

%\textit{introduction}--
Modern synthesis techniques have transformed atomic-scale materials into a versatile platform for exploring quantum phenomena \cite{hwang_review}.
A prominent example is the interface between the insulating oxides LaAlO$_3$ and SrTiO$_3$, where a high-mobility electron gas exhibiting superconductivity emerges \cite{hwang,sto_sc,PhysRevLett.104.126802,Bert2011,PhysRevLett.108.117003,PhysRevB.96.020504}.
Two-dimensional (2D) superconductivity has been extensively investigated in various systems, such as monolayer FeSe/SrTiO$_3$ \cite{FeSe/STO}, LaAlO$_3$/KTaO$_3$ \cite{doi:10.1021/acsami.7b12814,kto1,kto2}, EuO/KTaO$_3$ \cite{PhysRevLett.121.116803,PhysRevB.104.184505,kto1,Hua2022,Liu2023,Hua2024}, twisted bilayer graphene \cite{bilayer-graphene} and many other 2D materials \cite{2d-SC-review}. 
 Unlike their three-dimensional counterparts, the superconducting transition in 2D systems is governed by the Berezinskii-Kosterlitz-Thouless (BKT) mechanism \cite{Berezinsky:1970fr,Kosterlitz_1973,Kosterlitz_1974}.
Although proposed five decades ago, the implications of the BKT transition for 2D superconductors remain relatively underexplored \cite{Halperin1979}.

\begin{figure}[H]
	\begin{center}
		\fig{3.4in}{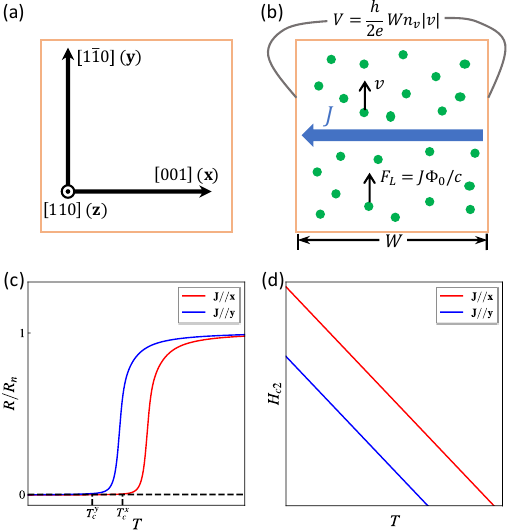}
		\caption{(a) Illustration of the three crystallographic directions of KTaO$_3$, labeled as $\mathbf{x}$, $\mathbf{y}$, and $\mathbf{z}$. (b) Schematic of the mechanism of vortex-induced finite resistance. The large blue arrow indicates the applied current $J$ through the sample, while the small green dots represent vortices. The current exerts a Lorentz force on each vortex, given by $F_L=J\Phi_0/c$, where $\Phi_0=hc/2e$ is the flux quantum and $c$ is the light speed. The vortex moves with velocity $v$, and the resulting voltage across the lateral width $W$ of the sample is $V=\frac{h}{2e}Wn_v|v|$, with $n_v$ denoting the vortex density. (c) Schematic illustration of the experimentally observed anisotropic critical temperature phenomenon. $T_c^x$ and $T_c^y$ are the critical temperatures at which zero-resistance state breaks down for currents along the $\mathbf{x}$- and $\mathbf{y}$-directions, respectively. (d) Schematic illustration of the observed anisotropic upper critical field $H_{c2}$ phenomenon. 
        \label{fig1}}
	\end{center}
% 		\vskip-1.5cm
\end{figure}

Recently, anisotropic superconducting behavior has been observed at the interface of EuO/KTaO$_3$(110) \cite{Hua2024}. Specifically,
different critical temperatures $T_c$ are reported when the current is applied along two orthogonal directions, $[001]$ and $[1\bar{1}0]$.
For further convenience, we designate the three crystallographic directions of KTaO$_3$--$[001]$, $[1\bar{1}0]$ and $[110]$--as $\mathbf{x}$, $\mathbf{y}$ and $\mathbf{z}$, respectively, as shown in Fig.~\ref{fig1}(a). 
The KTaO$_3$ sample exhibits two distinct directional $R-T$ curves in the superconducting transition region, schematically illustrated in Fig.~\ref{fig1}(c). 
If the critical temperature is defined as the point where resistance becomes nonzero, two critical temperatures are identified, with $T_c^x>T^y_c$.
This intriguing observation raises a fundamental question: Can the same system exhibit different superconducting transition temperatures?
Thermodynamically, this cannot be true for a global thermodynamic transition. Given that the BKT transition is governed by vortex-antivortex dynamics, we propose that anisotropic vortex motion underlies the observed anisotropic superconducting behavior, providing another route in addition to a recent theoretical proposal \cite{li2024theoryinfinitelyanisotropicphase}.

Another key observation supporting our proposal is the behavior of the superconducting transition under an external magnetic field $H$. In the presence of an external field, the superconductor-to-metal transition is primarily driven by vortex dynamics (in the absence of antivortices). Different meas-
\noindent
urement geometries are expected to yield different critical field values $H_{c2}$. This is indeed observed in KTaO$_3$ \cite{Hua2024}, as schematically illustrated in Fig.~\ref{fig1}(d).
The sample exhibits distinct $H_{c2}-T$ characteristics along the two directions, with $H_{c2}^x>H_{c2}^y$ over the entire temperature range. 
Therefore, these anisotropic transition behaviors are closely associated with the anisotropic motion of vortices.

% \begin{figure}[H]
% 	\begin{center}
% 		\fig{3.4in}{Fig1.pdf}
% 		\caption{(a) Three crystallographic directions of KTaO$_3$, denoted here as $\mathbf{x,y,\text{and } z}$. (b) Schematic diagram of the mechanism of vortex-induced finite resistance. The big blue arrow represents the current $J$ passing through the sample and small green dots are the vortices. The Magnus force exerted on the vortex by current is $F_M=J\Phi_0/c$, where $\Phi_0=hc/2e$ is the flux quantum and $c$ is the light speed. $v$ is the motion velocity of vortex. The lateral width of the sample is $W$. The voltage between two lateral sides is $V=\frac{h}{2e}Wn_v|v|$, where $n_v$ is the density of vortices. (c) Schematic illustration of the observed anisotropic critical temperature phenomenon. $T_c^x$ and $T_c^y$ are the critical temperatures at which zero-resistance state is broken. (d) Schematic illustration of the observed anisotropic upper critical field phenomenon. 
%         \label{fig1}}
% 	\end{center}
% % 		\vskip-1.5cm
% \end{figure}

%Vortex plays an important role for both cases. 
In the presence of a magnetic field, type-II superconductors allow the penetration of quantized vortices once the applied field exceeds the lower critical field $H_{c1}$, and this persists up to the upper critical field $H_{c2}$. In contrast, in the absence of an external field, 2D superconductors exhibit topological excitations in the form of vortex-antivortex pairs.
When vortices are present, superconductors develop finite resistance due to flux-flow dissipation\cite{tinkham1996introduction,kopnin2001theory,bardeen-stephen,Tinkham_PhysRevLett.13.804}. Intuitively, the mechanism of vortex-induced resistance is illustrated in Fig.~\ref{fig1}(b). An applied current $J$ exerts a Lorentz force on the vortex, driving its motion. If the vortex moves with velocity $v$, a voltage $V=\frac{h}{2e}Wn_v|v|$ develops across the lateral edges of the sample, in accordance with the Josephson relation \cite{Halperin1979,tinkham1996introduction}. This results in a resistivity of the form
\begin{equation}
    \rho\equiv (V/W)/J=\frac{h}{2e}n_v|v|/J.\label{eq:resistivity}
\end{equation}
where n$_v$ is the vortex density, and W is the sample width. 
\textit{Hence, under a fixed current $J$, one can determine the $\rho$ by finding $n_v$ and $v$, which is the strategy for the following discussions} \cite{Halperin1979}.
Spatial inhomogeneities introduce a pinning potential that modulates the free energy landscape experienced by vortices at different positions. This pinning potential can strongly influence vortex dynamics and, consequently, the resistance of the material. To capture the anisotropic vortex motion without loss of generality, we employ a model with an anisotropic pinning potential. In the $\mathrm{EuO/KTaO_3(110)}$ interface superconductor considered here \cite{kto1,Hua2024}, such an anisotropic pinning landscape may plausibly be associated with the stripe structure revealed by scanning superconducting quantum interference device (SQUID) measurements. Anisotropic pinning can also originate from impurities, grain boundaries, and other structural, fabrication-related, or geometric sources. Related vortex phenomena in anisotropic and other superconducting environments, including anisotropic melting-like behavior, smectic-like regimes, directional dynamical responses, and fast nonequilibrium vortex transport, have also been discussed in previous studies~\cite{PhysRevLett.90.087001,C.Reichhardt_2006,Guillamon2009,Guillamon2014,PhysRevB.93.014504,Dobrovolskiy2020}. More generally, the present problem may also be viewed in the broader context of topological-defect-driven two-dimensional melting and the associated KTHNY-like scenarios~\cite{Berezinskii:1972fet,Kosterlitz_1973,PhysRevLett.41.121,PhysRevB.19.2457}. In the following, we discuss two situations separately: the finite-field mixed-state case and the zero-field BKT case.

\begin{figure}[t]
	\begin{center}
		\fig{3.4in}{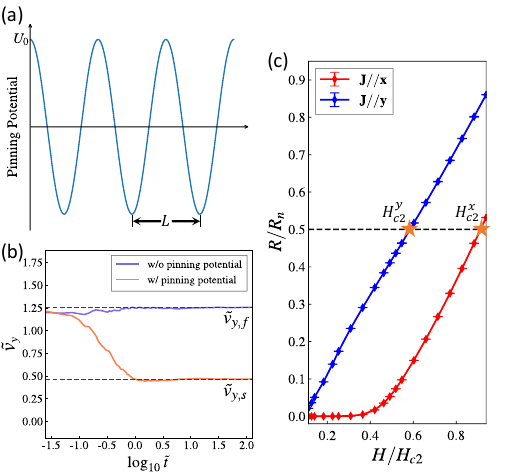}
		\caption{(a) Schematic illustration of the periodic pinning potential characterized by period $L$ and barrier height $U_0$. (b) Representative time evolution of $\tilde{v}_y$ at $H = 0.715H_{c2}$, where $H_{c2}=\Phi_0/(2\pi\xi^2)$,  when $\tilde{\mathbf{J}}$ is aligned with the $\mathbf{x}$-direction. The corresponding evolution without a pinning potential is also shown for comparison. Here, $\tilde{v}_{y,s}$ denotes the steady velocity after long time in the presence of pinning, while $\tilde{v}_{y,f}=2\pi\tilde{J}_x$ represents that without pinning. (c) $R-H$ curves obtained from dynamics simulations for two different directions of $\mathbf{J}$. The upper critical fields $H_{c2}^x$ and $H_{c2}^y$ are determined using the criterion of $50\%$ of the normal-state resistance. Simulation parameters for (b) and (c) are set as $\lambda = 80\,\xi$, $L = 8\,\xi$, $k_BT = 0.002\,(d\Phi_0^2/\lambda^2)$, $|\tilde{\mathbf{J}}| = 0.2$, $\tilde{U}_x = 3.5$, $\tilde{U}_y = 8.5$ and the sample size is $168\,\xi\times 168\,\xi$.\label{fig2}}
	\end{center}
		%\vskip-1.5cm
\end{figure}

\textit{Critical Field.}---
We begin by examining the finite-field mixed-state case, namely, the superconducting transition under a magnetic field at a fixed temperature. When the superconductor enters the mixed state, a vortex lattice forms. The vortex density is given by $n_v=B/\Phi_0$, where $B$ is the spatially averaged magnetic induction. For extreme type-II superconductors, $B$ is approximately equal to $H$ over a broad range of fields within the mixed state \cite{tinkham1996introduction}.
In addition to the Lorentz force, vortex motion is influenced by several other forces~\cite{tinkham1996introduction,PhysRevB.98.054510,zwanzig2001nonequilibrium,kopnin2001theory}. 
When a current flows through the system, it exerts a Lorentz force on the vortex, given by $\mathbf{F}_L=\mathbf{J}\times\mathbf{z}\Phi_0/c$, where $\mathbf{J}$ is the current density. %and $\mathbf{z}$ is the unit vector perpendicular to the sample plane.
Additionally, the vortex at position $\mathbf{r}$ experiences a viscous drag force $-\eta\, d \mathbf{r}/dt$ and a random fluctuation force $\sqrt{2\eta k_BT}\boldsymbol{\sigma}(t)$, where $T$ is the temperature and $\boldsymbol{\sigma}(t)$ is a white noise term. The viscous coefficient is related to the normal core of the vortex. $\eta = \Phi_0^2/(2\pi \xi^2 c^2\rho_n)$, where $\xi$ is the coherence length and $\rho_n$ is the resistivity in the normal state. 
Furthermore, the interaction force between vortices becomes significant, particularly at high vortex densities. This force is given by $\mathbf{F}_{vv}(\mathbf{r}_i-\mathbf{r}_j)=\frac{\Phi_0^2 d}{8\pi^2\lambda^3}K_1(r_{ij}/\lambda)\frac{\mathbf{r}_i-\mathbf{r}_j}{r_{ij}}$, where $d$ is the thickness of the sample film, $\lambda$ is the penetration length, $r_{ij}=|\mathbf{r}_i-\mathbf{r}_j|$, and $K_1$ is the first-order modified Bessel function of the second kind. 
We also emphasize the role of the pinning force. To simplify the analysis, we employ a periodic pinning potential in our model, as illustrated in Fig.~\ref{fig2}(a). This potential tends to trap vortices within its valleys, thereby reducing resistance. Specifically, the pinning potential in the two-dimensional model is expressed as
$U(\mathbf{r})=U_x\cos(\frac{2\pi}{L}x) + U_y\cos(\frac{2\pi}{L}y)$,
where $\mathbf{r}=(x,y)$, $L$ represents the periodic length, and $U_x$, $U_y$ are the barrier heights along the two respective directions, set to different values to introduce anisotropy. The neglect of transverse force terms is discussed in Sec.~IV of the Supplemental Material~\cite{Supplement}. By balancing all relevant forces, we arrive at the Langevin equation describing the motion of the vortex lattice
\begin{equation}
    \eta\frac{d\mathbf{r}_i}{dt} = \sum_{j\ne i}\mathbf{F}_{vv}(\mathbf{r}_i-\mathbf{r}_j) + \mathbf{F}_L + \sqrt{2\eta k_BT}\boldsymbol{\sigma}(t) -\nabla U.\label{eq:Langevin}
\end{equation}

Following the approach in Refs.~\cite{PhysRevLett.73.3580,PhysRevB.53.3520,PhysRevB.75.024506,PhysRevLett.98.267002,PhysRevB.84.104528,Dobramysl2013,PhysRevE.92.052124,Assi01112016}, the Langevin equation~\eqref{eq:Langevin} can be numerically simulated to capture the vortex dynamics. For convenience, we nondimensionalize the equation prior to simulation. We define $\mathbf{r}_i \equiv \xi\, \tilde{\mathbf{r}}_i$, $t \equiv \frac{\lambda^3}{d\xi c^2 \rho_n}\,\tilde{t}$,
$\mathbf{J} \equiv \frac{dc\Phi_0}{\lambda^3}\,\tilde{\mathbf{J}}$. With $\tilde{t}$ discretized, Eq.~\eqref{eq:Langevin} becomes
\begin{multline}
    \frac{\Delta \tilde{\mathbf{r}}_i}{\Delta \tilde{t}} = \frac{1}{4\pi}\sum_{j\ne i} K_1\left(\frac{\tilde{r}_{ij}}{\kappa}\right)\frac{\tilde{\mathbf{r}}_i-\tilde{\mathbf{r}}_j}{\tilde{r}_{ij}} \\+ 2\pi\, \tilde{\mathbf{J}}\times \mathbf{z}
    + 2\sqrt{\frac{\pi k_BT\lambda^3}{d\xi\Phi_0^2\Delta\tilde{t}}}\,\boldsymbol{\sigma}\left(\tilde{t}\right) + \tilde{\mathbf{F}}_{pin},\label{eq:Langevinnondim}
\end{multline}
where $\kappa=\lambda/\xi$ is the Ginzburg-Landau parameter and $
    \tilde{\mathbf{F}}_{pin} = \left(\tilde{U}_x\sin\left(\frac{2\pi}{L/\xi}\tilde{x}\right), \tilde{U}_y\sin\left(\frac{2\pi}{L/\xi}\tilde{y}\right)\right)$.
Assuming $\tilde{\mathbf{J}}$ is oriented along the $\mathbf{x}$-axis, we simulate Eq.~\eqref{eq:Langevinnondim} to track the time evolution of the vortex velocity. A representative evolution of $\tilde{v}_y$ is shown in Fig.~\ref{fig2}(b). As evident, $\tilde{v}_y$  eventually stabilizes at a steady value $\tilde{v}_{y,s}$, which is significantly suppressed compared to $\tilde{v}_{y,f}$, the corresponding steady velocity in the absence of pinning. This suppression arises from the hindrance of vortex motion caused by the pinning potential. By defining $n_v\equiv\tilde{n}_v/\xi^2$ and using Eq.~\eqref{eq:resistivity}, the resistance is expressed as:
 \begin{equation}
    R/R_n = \tilde{n}_v|\tilde{v}|/\tilde{J}.\label{eq:resistivitynondim}
\end{equation}
    
%Assuming $\tilde{\mathbf{J}}$ parallel to $\mathbf{x}$ and putting Eq.\eqref{eq:Langevinnondim} into simulation, we get the velocity evolution with time. Fig.\ref{fig2}(b) shows a typical evolution of $\tilde{v}_y$. It can be seen that $\tilde{v}_y$ eventually grows steady at $\tilde{v}_{y,s}$, which is suppressed compared to $\tilde{v}_{y,f}$, the steady velocity if there is no pinning potential. This suppression is obviously due to the obstruction of vortex motion by the pinning potential. After further defining $n_v\equiv\tilde{n}_v/\xi^2$ and taking advantage of Eq.\eqref{eq:resistivity}, the resistivity is calculated as
%\begin{equation}
%    \rho/\rho_n = \tilde{n}_v|\tilde{v}|/\tilde{J}.\label{eq:resistivitynondim}
%\end{equation}

Applying this procedure across various values of $H$ and for both orientations of $\tilde{\mathbf{J}}$, we obtain the $R-H$ curves shown in Fig.~\ref{fig2}(c). When $\tilde{U}_x\ll\tilde{U}_y$, the resistance along the $\mathbf{y}$-direction is only marginally reduced, whereas that along the $\mathbf{x}$-direction experiences significant suppression at low $H$. However, as $H$ increases, the suppression diminishes and eventually disappears. This behavior along the $\mathbf{x}$-direction is intuitive: at low $H$, vortices are sparsely distributed, making them more easily trapped within the valleys of the pinning potential. As $H$ increases, the vortex density rises. Clusters of vortices collectively confront the pinning barriers and eventually overcome them. As a result, the influence of the pinning potential becomes progressively weaker, thereby reducing its impact on the resistance. Using a criterion of $50\%$ of the normal-state resistance as in Ref.~\cite{Hua2024}, we identify two distinct upper critical fields, $H_{c2}^x>H_{c2}^y$, corresponding to the $\mathbf{x}$- and $\mathbf{y}$-directions, respectively. This model can be extended to all temperatures below $T_c$, yielding the results shown in Fig.~\ref{fig1}(d).

%Operating this procedure for different $H$ and for both directions of $\tilde{\mathbf{J}}$, we get the $R-H$ curves in Fig.\ref{fig2}(c). With $\tilde{U}_x\ll\tilde{U}_y$, the resistivity along $\mathbf{y}$ direction is only slightly reduced, while that along $\mathbf{x}$ direction is prominently suppressed when $H$ is small but eventually escapes the suppression as $H$ gets large. The behavior along $\mathbf{x}$ direction is reasonable. When $H$ is small, vortices have a sparse distribution so that they are easy to be pinned in the valley of pinning potential. As $H$ increases, on the other hand, a cluster of vortices faces the barrier together and eventually they will overcome it. The influence of pinning potential becomes less and less important and the reduction of resistivity by it gets smaller and smaller. By a criterion of $50\%$ normal-state resistance, two different upper critical fields are determined for $\mathbf{x}$ and $\mathbf{y}$ directions with $H_{c2}^x>H_{c2}^y$. This model can be applied to all temperatures below $T_c$ to produce Fig.\ref{fig1}(d).

\textit{BKT.}---
We now turn to the zero-field case, namely, the superconducting transition crossing temperature.
It is well established that 2D superconductors undergo the BKT transition \cite{Kosterlitz_1974,Halperin1979}. In this zero-field BKT case, vortices are not introduced by an applied magnetic field. Instead, the relevant excitations are thermally excited vortex-antivortex pairs. As the temperature exceeds the critical $T_c$, vortex-antivortex pairs unbind into free vortices and antivortices, which destroy phase coherence and lead to finite resistance. This resistance emerges through the same mechanism as in the presence of a magnetic field \cite{Halperin1979}. Below $T_c$, vortex-antivortex pairs remain bound, preserving phase coherence. However, due to finite-current effects, the resistance is not exactly zero. When a current flows through the sample, as discussed earlier, it exerts Lorentz forces on both the vortex and the antivortex; however, due to their opposite topological charges, these forces act in opposite directions. The opposite forces tend to pull the bound pairs apart. Upon unbinding, free vortices and antivortices are produced, leading to energy dissipation similar to the behavior above $T_c$.

\begin{figure}[t]
	\begin{center}
		\fig{3.4in}{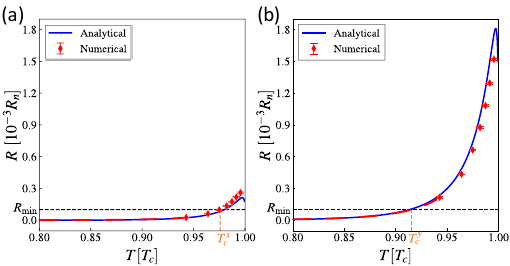}
		\caption{Analytical and numerical $R$-$T$ curves in the zero-field BKT case. (a) $\mathbf{J//x}$. (b) $\mathbf{J//y}$. In each panel, the blue solid curve is the analytical result obtained from Eq.~\eqref{eq:resistivity_pin}, the red symbols are the numerical results from the full 2D Langevin simulation, and the red dashed curve is the low-temperature estimate obtained by substituting the velocity $|\tilde{v}|$, computed at the lowest numerically tractable temperature (where only one vortex and one antivortex are generated), into Eq.~\eqref{eq:resistivitynondim}. $R_{\mathrm{min}}$ denotes the minimum detectable resistance in experiments. $T_c^x$ and $T_c^y$ are the experimentally determined critical temperatures along the $\mathbf{x}$- and $\mathbf{y}$-directions, respectively, while $T_c$ denotes the theoretical BKT transition temperature. Parameters for both analytical and numerical cases are $b = 1$, $\tau_c = 0.15$, $k_BT_c = 0.008\,(d\Phi_0^2/\lambda^2)$, $J/J_0=0.1$, $U_x=1.1\,k_BT_c$ and $U_y=2.7\,k_BT_c$ (corresponding to $|\tilde{\mathbf{J}}|=0.2$, $\tilde{U}_x=3.5$, and $\tilde{U}_y=8.5$ in the previous section). The same spatial parameters as those in Fig.~\ref{fig2} are used for the numerical calculations.\label{fig3}}
	\end{center}
		%\vskip-1.5cm
\end{figure}

We first discuss the situation without a pinning potential at $T<T_c$. 
As shown in Fig.~\ref{fig2}(b), the long-time steady velocity of vortices $\tilde{v}_{y,f} = 2\pi \tilde{J}_x$, giving $R/R_n = 2\pi n_v\xi^2$,
where $n_v$ denotes the density of current-induced vortices and antivortices. The dependence of $n_v$ on temperature and current has been investigated in Refs.~\cite{PhysRevLett.40.783,PhysRevLett.40.780,PhysRevB.18.3204,Halperin1979}. The resulting resistance is given\cite{Halperin1979} as
\begin{subequations}
\begin{equation}
    R/R_n = s(J/J_0)^{2+s/2}, \label{eq:resistivity_nopin}
\end{equation}
\begin{equation}
    J_0 \equiv T_ce/\hbar\xi,
\end{equation}
\end{subequations}
where $s \equiv \text{max}[s_T,s_J]$ with $s_T \equiv 2\pi|\tau/b\tau_c|^{1/2}, s_J \equiv 1/\ln(J_0/J)$. Here, $b$ is a dimensionless constant of order unity, $\tau \equiv (T-T_c)/T_c $ is the reduced temperature, and $\tau_c \equiv (T^0_c-T_c)/T_c$, where $T_c^0$ is the BCS critical temperature.

Next, we address the case with a pinning potential. The situation becomes simpler in the BKT case, as the density of current-induced vortices and antivortices remains low when the current is small. Consequently, the interaction between vortices can be safely neglected. To fully leverage analytical tractability, we study the one-dimensional (1D) Langevin equation
\begin{equation}
    \eta\,\frac{dx}{dt} = F_L + \sqrt{2\eta k_BT}\sigma(t) -\frac{\partial U}{\partial x}
\end{equation}
where $U(x) = U_0\cos(\frac{2\pi}{L}x)$. An analytical solution\cite{PhysRevLett.87.010602,PhysRevB.98.054510} for the long-time steady velocity exists, which reads, to first-order in $F_L$,
    \begin{equation}
    v = \frac{L}{\beta \eta}\frac{\beta L}{\int_0^Ldy\,\mathcal{I}(y)}F_L + O\left(F_L^2\right),
\end{equation}
where $\mathcal{I}(y) = \int_0^Ldx\,e^{\beta[U(x)-U(x-y)]} $ with $\beta=1/k_BT$.
With our specific $U(x)$, the resistance is derived to be
\begin{equation}
    R/R_n = 2\pi n_v\xi^2/[I_0(\beta U_0)]^2,
\end{equation}
where $I_0$ is the zeroth-order modified Bessel function of the first kind. Obviously, the pinning potential reduces the resistance by a factor of $1/[I_0(\beta U_0)]^2$. Thus, the explicit expression for the resistance below $T_c$ in the presence of a cosine-wave pinning potential is
\begin{equation}
    R/R_n = s(J/J_0)^{2+s/2}/[I_0(\beta U_0)]^2.\label{eq:resistivity_pin}
\end{equation}

If we identify $U_0$ with $U_x$ and $U_y$ in the 2D model discussed in the previous section, Eq.~\eqref{eq:resistivity_pin} with two different values of $U_0$ yields the resistance for $\mathbf{J//x}$ and $\mathbf{J//y}$ respectively. To validate the applicability of the 1D analytical results to the 2D case, we also perform a full 2D numerical simulation including the vortex-antivortex interactions. The computation details are provided in Sec.~I of the Supplemental Material~\cite{Supplement}. The analytical and numerical $R$-$T$ curves are compared directly in Fig.~\ref{fig3}, with panel (a) for $\mathbf{J//x}$ and panel (b) for $\mathbf{J//y}$. As evident from the figure, the analytical and numerical results agree closely. In experiments, resistance is typically determined by measuring the voltage under a fixed probing current. Due to the finite resolution of voltage measurements, there exists a minimum detectable resistance, $R_{\mathrm{min}}$, which effectively serves as the operational definition of zero resistance \cite{li2024theoryinfinitelyanisotropicphase}. As previously discussed, the resistance below $T_c$ is not strictly zero. When it exceeds $R_{\mathrm{min}}$, the corresponding temperature is identified as the experimental critical temperature, which is lower than the theoretical BKT transition temperature $T_c$. The anisotropic pinning potential strongly suppresses resistance in the $\mathbf{x}$-direction, causing it to exceed $R_{\mathrm{min}}$ at a higher temperature. Consequently, the experimentally observed critical temperature for the $\mathbf{x}$-direction appears elevated. Fig.~\ref{fig3} effectively captures the dominant behavior in the transition region depicted in Fig.~\ref{fig1}(c). It is worth noting that the slight dip in the analytical curves near $T_c$ in Fig.~\ref{fig3} is not of practical importance, as this region falls outside the applicable range of Eq.~\eqref{eq:resistivity_nopin}. Moreover, once the temperature exceeds $T_c$, free vortices emerge spontaneously, leading to a pronounced rise in resistance.

\begin{figure}[t]
	\begin{center}
		\fig{3.4in}{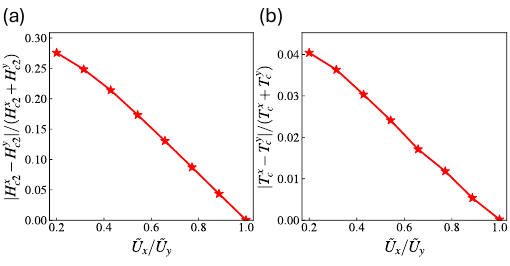}
		\caption{Dependence of the directional anisotropy on the pinning-potential ratio $\tilde{U}_x/\tilde{U}_y$. (a) $|H_{c2}^{x}-H_{c2}^{y}|/(H_{c2}^{x}+H_{c2}^{y})$ in the finite-field mixed-state regime. (b) $|T_{c}^{x}-T_{c}^{y}|/(T_{c}^{x}+T_{c}^{y})$ in the zero-field BKT regime. In both cases, $\tilde{U}_y$ is fixed and $\tilde{U}_x$ is varied. Other parameters are the same as those in Fig.~\ref{fig2} for (a) and Fig.~\ref{fig3} for (b), except for the variation of $\tilde{U}_x$.\label{fig4}}
	\end{center}
		%\vskip-1.5cm
\end{figure}

To further clarify the role of the pinning anisotropy, we also perform systematic scans of $\tilde{U}_x/\tilde{U}_y$ by fixing $\tilde{U}_y$ and varying $\tilde{U}_x$ in both the finite-field mixed-state and zero-field BKT regimes. The resulting anisotropy measures, $|H_{c2}^{x}-H_{c2}^{y}|/(H_{c2}^{x}+H_{c2}^{y})$ and $|T_{c}^{x}-T_{c}^{y}|/(T_{c}^{x}+T_{c}^{y})$, are shown in Fig.~\ref{fig4}(a) and Fig.~\ref{fig4}(b), respectively. In both cases, the anisotropy decreases monotonically as $\tilde{U}_x/\tilde{U}_y$ approaches $1$, and becomes nearly zero in the isotropic limit. These results further support the interpretation that the directional differences in the critical temperatures and critical fields originate from the anisotropic pinning potential.

%\textit{Discussion and summary.}---
%In summary, the anisotropic pinning potential can significantly modify the vortex motion efficiency under different current direction. And this feature is general both in BKT vortex-antivortex driven motion and mixed state vortex motion.
%One unique question is the origin of the pinning potential.
%In KTaO$_3$ interface superconductor \cite{kto1,Hua2024}, this pinning potential most likely comes from the stripe structure, as probed by scanning superconducting quantum interference device (SQUID) measurements. This stripe structure can originate from the  ferromagnetic EuO substrate and other structure anisotropy. On the other hand, we want to point out the anisotropic pinning potential is general, which can come from impurities, grain boundaries etc., are inevitable in the fabrication process and geometry design. Therefore, anisotropic pinning potential is a natural candidate for the explanation of experimental observations.
%In this work, our approach is still based on the semiclassical Langevin equation. We call for further microscopic theoretical investigation at microscopic level. We hope our theory and findings provide a new insight toward the two-dimensional superconductor  anisotropy.

\textit{Summary and Discussion.}---
In summary, the anisotropic pinning potential can substantially affect the efficiency of vortex motion, depending on the direction of the applied current. This characteristic is applicable to both the BKT vortex-antivortex motion and the mixed-state vortex motion. As noted above, in the case of the KTaO$_3$ interface superconductor \cite{kto1,Hua2024}, the pinning potential is likely attributed to the stripe structure, as revealed by scanning superconducting quantum interference device (SQUID) measurements. This stripe structure may arise from the ferromagnetic EuO substrate or other structural anisotropies. Therefore, such an anisotropic pinning potential provides a natural explanation for the anisotropic superconducting transition behaviors discussed above.

%Based on our proposal, the current pays an important role driven vortex motion. Hence, the probing current also influences the 
%critical temperature anisotropy. We find that the difference between two critical temperatures $T_c^x$ and $T_c^y$ becomes larger as the probing current increases (see Supplemental Material). In addition, the $V-I$ characteristic at a fixed temperature near the detected $T_c$ is expected to be $V\propto I^\alpha$ with the same $\alpha$ ($\sim 3$) on both directions as the BKT transition theory predicts. Moreover, this anisotropic pinning potential may potentially be fabricated by modern quantum design techniques these days, thereby generally enabling the realization of multiple "critical temperatures" in two-dimensional superconductors. On the other hand,  our approach remains grounded in the semiclassical Langevin equation in this study. We call for further microscopic theoretical investigation at microscopic level to deepen the understanding of this phenomenon. We hope that our theoretical framework and findings offer new insights into the anisotropic behavior of two-dimensional superconductors.

Based on our proposal, the probing current plays a pivotal role in driving vortex motion and therefore also affects the transition anisotropy. Direct numerical calculations show that the normalized mixed-state critical field anisotropy, defined as $|H_{c2}^x-H_{c2}^y|/(H_{c2}^x+H_{c2}^y)$, decreases monotonically but only mildly with increasing probing current, whereas the normalized BKT critical temperature anisotropy, defined as $|T_c^x-T_c^y|/(T_c^x+T_c^y)$, increases monotonically. For the simplified analytical discussion of the BKT case and the direct numerical results in both regimes, see Secs.~II and III of the Supplemental Material~\cite{Supplement}. Furthermore, advancements in modern quantum design techniques may enable the fabrication of such an anisotropic pinning potential, potentially allowing the realization of multiple ``critical temperatures" in two-dimensional superconductors. However, our analysis in this study is grounded in a semiclassical Langevin framework. To deepen our understanding of this phenomenon, further investigations at the microscopic level are necessary. We hope that our theoretical approach and findings offer insights into the anisotropic behavior of two-dimensional superconductors.

\textit{Acknowledgement:}
We thank Ziji Xiang and Zixiang Li for their thoughtful discussions. We acknowledge the support by the Ministry of Science and Technology (Grant No. 2022YFA1403900) and the National Natural Science Foundation of China (Grant NSFC-12494594), the Chinese Academy of Sciences Project for Young Scientists in Basic Research (2022YSBR-048), the New Cornerstone Investigator Program.

%\bibliographystyle{naturemag}
% \bibliography{reference}

\begin{thebibliography}{51}%
\makeatletter
\providecommand \@ifxundefined [1]{%
 \@ifx{#1\undefined}
}%
\providecommand \@ifnum [1]{%
 \ifnum #1\expandafter \@firstoftwo
 \else \expandafter \@secondoftwo
 \fi
}%
\providecommand \@ifx [1]{%
 \ifx #1\expandafter \@firstoftwo
 \else \expandafter \@secondoftwo
 \fi
}%
\providecommand \natexlab [1]{#1}%
\providecommand \enquote  [1]{``#1''}%
\providecommand \bibnamefont  [1]{#1}%
\providecommand \bibfnamefont [1]{#1}%
\providecommand \citenamefont [1]{#1}%
\providecommand \href@noop [0]{\@secondoftwo}%
\providecommand \href [0]{\begingroup \@sanitize@url \@href}%
\providecommand \@href[1]{\@@startlink{#1}\@@href}%
\providecommand \@@href[1]{\endgroup#1\@@endlink}%
\providecommand \@sanitize@url [0]{\catcode `\\12\catcode `\$12\catcode `\&12\catcode `\#12\catcode `\^12\catcode `\_12\catcode `\%12\relax}%
\providecommand \@@startlink[1]{}%
\providecommand \@@endlink[0]{}%
\providecommand \url  [0]{\begingroup\@sanitize@url \@url }%
\providecommand \@url [1]{\endgroup\@href {#1}{\urlprefix }}%
\providecommand \urlprefix  [0]{URL }%
\providecommand \Eprint [0]{\href }%
\providecommand \doibase [0]{http://dx.doi.org/}%
\providecommand \selectlanguage [0]{\@gobble}%
\providecommand \bibinfo  [0]{\@secondoftwo}%
\providecommand \bibfield  [0]{\@secondoftwo}%
\providecommand \translation [1]{[#1]}%
\providecommand \BibitemOpen [0]{}%
\providecommand \bibitemStop [0]{}%
\providecommand \bibitemNoStop [0]{.\EOS\space}%
\providecommand \EOS [0]{\spacefactor3000\relax}%
\providecommand \BibitemShut  [1]{\csname bibitem#1\endcsname}%
\let\auto@bib@innerbib\@empty
%</preamble>
\bibitem [{\citenamefont {Hwang}\ \emph {et~al.}(2012)\citenamefont {Hwang}, \citenamefont {Iwasa}, \citenamefont {Kawasaki}, \citenamefont {Keimer}, \citenamefont {Nagaosa},\ and\ \citenamefont {Tokura}}]{hwang_review}%
  \BibitemOpen
  \bibfield  {author} {\bibinfo {author} {\bibfnamefont {H.~Y.}\ \bibnamefont {Hwang}}, \bibinfo {author} {\bibfnamefont {Y.}~\bibnamefont {Iwasa}}, \bibinfo {author} {\bibfnamefont {M.}~\bibnamefont {Kawasaki}}, \bibinfo {author} {\bibfnamefont {B.}~\bibnamefont {Keimer}}, \bibinfo {author} {\bibfnamefont {N.}~\bibnamefont {Nagaosa}}, \ and\ \bibinfo {author} {\bibfnamefont {Y.}~\bibnamefont {Tokura}},\ }\bibfield  {title} {\enquote {\bibinfo {title} {Emergent phenomena at oxide interfaces},}\ }\href {\doibase 10.1038/nmat3223} {\bibfield  {journal} {\bibinfo  {journal} {Nature Materials}\ }\textbf {\bibinfo {volume} {11}},\ \bibinfo {pages} {103--113} (\bibinfo {year} {2012})}\BibitemShut {NoStop}%
\bibitem [{\citenamefont {Ohtomo}\ and\ \citenamefont {Hwang}(2004)}]{hwang}%
  \BibitemOpen
  \bibfield  {author} {\bibinfo {author} {\bibfnamefont {A.}~\bibnamefont {Ohtomo}}\ and\ \bibinfo {author} {\bibfnamefont {H.~Y.}\ \bibnamefont {Hwang}},\ }\bibfield  {title} {\enquote {\bibinfo {title} {A high-mobility electron gas at the $\mathrm{LaAlO_3/SrTiO_3}$ heterointerface},}\ }\href {\doibase 10.1038/nature02308} {\bibfield  {journal} {\bibinfo  {journal} {Nature}\ }\textbf {\bibinfo {volume} {427}},\ \bibinfo {pages} {423--426} (\bibinfo {year} {2004})}\BibitemShut {NoStop}%
\bibitem [{\citenamefont {Reyren}\ \emph {et~al.}(2007)\citenamefont {Reyren}, \citenamefont {Thiel}, \citenamefont {Caviglia}, \citenamefont {Kourkoutis}, \citenamefont {Hammerl}, \citenamefont {Richter}, \citenamefont {Schneider}, \citenamefont {Kopp}, \citenamefont {Rüetschi}, \citenamefont {Jaccard}, \citenamefont {Gabay}, \citenamefont {Muller}, \citenamefont {Triscone},\ and\ \citenamefont {Mannhart}}]{sto_sc}%
  \BibitemOpen
  \bibfield  {author} {\bibinfo {author} {\bibfnamefont {N.}~\bibnamefont {Reyren}}, \bibinfo {author} {\bibfnamefont {S.}~\bibnamefont {Thiel}}, \bibinfo {author} {\bibfnamefont {A.~D.}\ \bibnamefont {Caviglia}}, \bibinfo {author} {\bibfnamefont {L.~Fitting}\ \bibnamefont {Kourkoutis}}, \bibinfo {author} {\bibfnamefont {G.}~\bibnamefont {Hammerl}}, \bibinfo {author} {\bibfnamefont {C.}~\bibnamefont {Richter}}, \bibinfo {author} {\bibfnamefont {C.~W.}\ \bibnamefont {Schneider}}, \bibinfo {author} {\bibfnamefont {T.}~\bibnamefont {Kopp}}, \bibinfo {author} {\bibfnamefont {A.-S.}\ \bibnamefont {Rüetschi}}, \bibinfo {author} {\bibfnamefont {D.}~\bibnamefont {Jaccard}}, \bibinfo {author} {\bibfnamefont {M.}~\bibnamefont {Gabay}}, \bibinfo {author} {\bibfnamefont {D.~A.}\ \bibnamefont {Muller}}, \bibinfo {author} {\bibfnamefont {J.-M.}\ \bibnamefont {Triscone}}, \ and\ \bibinfo {author} {\bibfnamefont {J.}~\bibnamefont {Mannhart}},\ }\bibfield  {title} {\enquote {\bibinfo {title} {Superconducting interfaces
  between insulating oxides},}\ }\href {\doibase 10.1126/science.1146006} {\bibfield  {journal} {\bibinfo  {journal} {Science}\ }\textbf {\bibinfo {volume} {317}},\ \bibinfo {pages} {1196--1199} (\bibinfo {year} {2007})}\BibitemShut {NoStop}%
\bibitem [{\citenamefont {Ben~Shalom}\ \emph {et~al.}(2010)\citenamefont {Ben~Shalom}, \citenamefont {Sachs}, \citenamefont {Rakhmilevitch}, \citenamefont {Palevski},\ and\ \citenamefont {Dagan}}]{PhysRevLett.104.126802}%
  \BibitemOpen
  \bibfield  {author} {\bibinfo {author} {\bibfnamefont {M.}~\bibnamefont {Ben~Shalom}}, \bibinfo {author} {\bibfnamefont {M.}~\bibnamefont {Sachs}}, \bibinfo {author} {\bibfnamefont {D.}~\bibnamefont {Rakhmilevitch}}, \bibinfo {author} {\bibfnamefont {A.}~\bibnamefont {Palevski}}, \ and\ \bibinfo {author} {\bibfnamefont {Y.}~\bibnamefont {Dagan}},\ }\bibfield  {title} {\enquote {\bibinfo {title} {Tuning spin-orbit coupling and superconductivity at the $\mathrm{SrTiO_3/LaAlO_3}$ interface: A magnetotransport study},}\ }\href {\doibase 10.1103/PhysRevLett.104.126802} {\bibfield  {journal} {\bibinfo  {journal} {Phys. Rev. Lett.}\ }\textbf {\bibinfo {volume} {104}},\ \bibinfo {pages} {126802} (\bibinfo {year} {2010})}\BibitemShut {NoStop}%
\bibitem [{\citenamefont {Bert}\ \emph {et~al.}(2011)\citenamefont {Bert}, \citenamefont {Kalisky}, \citenamefont {Bell}, \citenamefont {Kim}, \citenamefont {Hikita}, \citenamefont {Hwang},\ and\ \citenamefont {Moler}}]{Bert2011}%
  \BibitemOpen
  \bibfield  {author} {\bibinfo {author} {\bibfnamefont {Julie~A.}\ \bibnamefont {Bert}}, \bibinfo {author} {\bibfnamefont {Beena}\ \bibnamefont {Kalisky}}, \bibinfo {author} {\bibfnamefont {Christopher}\ \bibnamefont {Bell}}, \bibinfo {author} {\bibfnamefont {Minu}\ \bibnamefont {Kim}}, \bibinfo {author} {\bibfnamefont {Yasuyuki}\ \bibnamefont {Hikita}}, \bibinfo {author} {\bibfnamefont {Harold~Y.}\ \bibnamefont {Hwang}}, \ and\ \bibinfo {author} {\bibfnamefont {Kathryn~A.}\ \bibnamefont {Moler}},\ }\bibfield  {title} {\enquote {\bibinfo {title} {Direct imaging of the coexistence of ferromagnetism and superconductivity at the $\mathrm{LaAlO_3/SrTiO_3}$ interface},}\ }\href {\doibase 10.1038/nphys2079} {\bibfield  {journal} {\bibinfo  {journal} {Nature Physics}\ }\textbf {\bibinfo {volume} {7}},\ \bibinfo {pages} {767--771} (\bibinfo {year} {2011})}\BibitemShut {NoStop}%
\bibitem [{\citenamefont {Michaeli}\ \emph {et~al.}(2012)\citenamefont {Michaeli}, \citenamefont {Potter},\ and\ \citenamefont {Lee}}]{PhysRevLett.108.117003}%
  \BibitemOpen
  \bibfield  {author} {\bibinfo {author} {\bibfnamefont {Karen}\ \bibnamefont {Michaeli}}, \bibinfo {author} {\bibfnamefont {Andrew~C.}\ \bibnamefont {Potter}}, \ and\ \bibinfo {author} {\bibfnamefont {Patrick~A.}\ \bibnamefont {Lee}},\ }\bibfield  {title} {\enquote {\bibinfo {title} {Superconducting and ferromagnetic phases in $\mathrm{SrTiO_3/LaAlO_3}$ oxide interface structures: Possibility of finite momentum pairing},}\ }\href {\doibase 10.1103/PhysRevLett.108.117003} {\bibfield  {journal} {\bibinfo  {journal} {Phys. Rev. Lett.}\ }\textbf {\bibinfo {volume} {108}},\ \bibinfo {pages} {117003} (\bibinfo {year} {2012})}\BibitemShut {NoStop}%
\bibitem [{\citenamefont {Monteiro}\ \emph {et~al.}(2017)\citenamefont {Monteiro}, \citenamefont {Groenendijk}, \citenamefont {Groen}, \citenamefont {de~Bruijckere}, \citenamefont {Gaudenzi}, \citenamefont {van~der Zant},\ and\ \citenamefont {Caviglia}}]{PhysRevB.96.020504}%
  \BibitemOpen
  \bibfield  {author} {\bibinfo {author} {\bibfnamefont {A.~M. R. V.~L.}\ \bibnamefont {Monteiro}}, \bibinfo {author} {\bibfnamefont {D.~J.}\ \bibnamefont {Groenendijk}}, \bibinfo {author} {\bibfnamefont {I.}~\bibnamefont {Groen}}, \bibinfo {author} {\bibfnamefont {J.}~\bibnamefont {de~Bruijckere}}, \bibinfo {author} {\bibfnamefont {R.}~\bibnamefont {Gaudenzi}}, \bibinfo {author} {\bibfnamefont {H.~S.~J.}\ \bibnamefont {van~der Zant}}, \ and\ \bibinfo {author} {\bibfnamefont {A.~D.}\ \bibnamefont {Caviglia}},\ }\bibfield  {title} {\enquote {\bibinfo {title} {Two-dimensional superconductivity at the (111)$\mathrm{LaAlO}{}_{3}/\mathrm{SrTiO}{}_{3}$ interface},}\ }\href {\doibase 10.1103/PhysRevB.96.020504} {\bibfield  {journal} {\bibinfo  {journal} {Phys. Rev. B}\ }\textbf {\bibinfo {volume} {96}},\ \bibinfo {pages} {020504} (\bibinfo {year} {2017})}\BibitemShut {NoStop}%
\bibitem [{\citenamefont {Qing-Yan}\ \emph {et~al.}(2012)\citenamefont {Qing-Yan}, \citenamefont {Zhi}, \citenamefont {Wen-Hao}, \citenamefont {Zuo-Cheng}, \citenamefont {Jin-Song}, \citenamefont {Wei}, \citenamefont {Hao}, \citenamefont {Yun-Bo}, \citenamefont {Peng}, \citenamefont {Kai}, \citenamefont {Jing}, \citenamefont {Can-Li}, \citenamefont {Ke}, \citenamefont {Jin-Feng}, \citenamefont {Shuai-Hua}, \citenamefont {Ya-Yu}, \citenamefont {Li-Li}, \citenamefont {Xi}, \citenamefont {Xu-Cun},\ and\ \citenamefont {Qi-Kun}}]{FeSe/STO}%
  \BibitemOpen
  \bibfield  {author} {\bibinfo {author} {\bibfnamefont {Wang}\ \bibnamefont {Qing-Yan}}, \bibinfo {author} {\bibfnamefont {Li}~\bibnamefont {Zhi}}, \bibinfo {author} {\bibfnamefont {Zhang}\ \bibnamefont {Wen-Hao}}, \bibinfo {author} {\bibfnamefont {Zhang}\ \bibnamefont {Zuo-Cheng}}, \bibinfo {author} {\bibfnamefont {Zhang}\ \bibnamefont {Jin-Song}}, \bibinfo {author} {\bibfnamefont {Li}~\bibnamefont {Wei}}, \bibinfo {author} {\bibfnamefont {Ding}\ \bibnamefont {Hao}}, \bibinfo {author} {\bibfnamefont {Ou}~\bibnamefont {Yun-Bo}}, \bibinfo {author} {\bibfnamefont {Deng}\ \bibnamefont {Peng}}, \bibinfo {author} {\bibfnamefont {Chang}\ \bibnamefont {Kai}}, \bibinfo {author} {\bibfnamefont {Wen}\ \bibnamefont {Jing}}, \bibinfo {author} {\bibfnamefont {Song}\ \bibnamefont {Can-Li}}, \bibinfo {author} {\bibfnamefont {He}~\bibnamefont {Ke}}, \bibinfo {author} {\bibfnamefont {Jia}\ \bibnamefont {Jin-Feng}}, \bibinfo {author} {\bibfnamefont {Ji}~\bibnamefont {Shuai-Hua}}, \bibinfo {author} {\bibfnamefont {Wang}\
  \bibnamefont {Ya-Yu}}, \bibinfo {author} {\bibfnamefont {Wang}\ \bibnamefont {Li-Li}}, \bibinfo {author} {\bibfnamefont {Chen}\ \bibnamefont {Xi}}, \bibinfo {author} {\bibfnamefont {Ma}~\bibnamefont {Xu-Cun}}, \ and\ \bibinfo {author} {\bibfnamefont {Xue}\ \bibnamefont {Qi-Kun}},\ }\bibfield  {title} {\enquote {\bibinfo {title} {Interface-induced high-temperature superconductivity in single unit-cell {FeSe} films on {SrTiO$_3$}},}\ }\href {\doibase 10.1088/0256-307X/29/3/037402} {\bibfield  {journal} {\bibinfo  {journal} {Chin. Phys. Lett.}\ }\textbf {\bibinfo {volume} {29}},\ \bibinfo {pages} {037402--037402} (\bibinfo {year} {2012})}\BibitemShut {NoStop}%
\bibitem [{\citenamefont {Zhang}\ \emph {et~al.}(2017)\citenamefont {Zhang}, \citenamefont {Zhang}, \citenamefont {Yan}, \citenamefont {Zhang}, \citenamefont {Zhang}, \citenamefont {Zhang}, \citenamefont {Han}, \citenamefont {Gu}, \citenamefont {Liu}, \citenamefont {Chen}, \citenamefont {Shen},\ and\ \citenamefont {Sun}}]{doi:10.1021/acsami.7b12814}%
  \BibitemOpen
  \bibfield  {author} {\bibinfo {author} {\bibfnamefont {Hui}\ \bibnamefont {Zhang}}, \bibinfo {author} {\bibfnamefont {Hongrui}\ \bibnamefont {Zhang}}, \bibinfo {author} {\bibfnamefont {Xi}~\bibnamefont {Yan}}, \bibinfo {author} {\bibfnamefont {Xuejing}\ \bibnamefont {Zhang}}, \bibinfo {author} {\bibfnamefont {Qinghua}\ \bibnamefont {Zhang}}, \bibinfo {author} {\bibfnamefont {Jing}\ \bibnamefont {Zhang}}, \bibinfo {author} {\bibfnamefont {Furong}\ \bibnamefont {Han}}, \bibinfo {author} {\bibfnamefont {Lin}\ \bibnamefont {Gu}}, \bibinfo {author} {\bibfnamefont {Banggui}\ \bibnamefont {Liu}}, \bibinfo {author} {\bibfnamefont {Yuansha}\ \bibnamefont {Chen}}, \bibinfo {author} {\bibfnamefont {Baogen}\ \bibnamefont {Shen}}, \ and\ \bibinfo {author} {\bibfnamefont {Jirong}\ \bibnamefont {Sun}},\ }\bibfield  {title} {\enquote {\bibinfo {title} {Highly mobile two-dimensional electron gases with a strong gating effect at the amorphous $\mathrm{LaAlO_3/KTaO_3}$ interface},}\ }\href {\doibase 10.1021/acsami.7b12814}
  {\bibfield  {journal} {\bibinfo  {journal} {ACS Applied Materials \& Interfaces}\ }\textbf {\bibinfo {volume} {9}},\ \bibinfo {pages} {36456--36461} (\bibinfo {year} {2017})}\BibitemShut {NoStop}%
\bibitem [{\citenamefont {Liu}\ \emph {et~al.}(2021)\citenamefont {Liu}, \citenamefont {Yan}, \citenamefont {Jin}, \citenamefont {Ma}, \citenamefont {Hsiao}, \citenamefont {Lin}, \citenamefont {Bretz-Sullivan}, \citenamefont {Zhou}, \citenamefont {Pearson}, \citenamefont {Fisher}, \citenamefont {Jiang}, \citenamefont {Han}, \citenamefont {Zuo}, \citenamefont {Wen}, \citenamefont {Fong}, \citenamefont {Sun}, \citenamefont {Zhou},\ and\ \citenamefont {Bhattacharya}}]{kto1}%
  \BibitemOpen
  \bibfield  {author} {\bibinfo {author} {\bibfnamefont {Changjiang}\ \bibnamefont {Liu}}, \bibinfo {author} {\bibfnamefont {Xi}~\bibnamefont {Yan}}, \bibinfo {author} {\bibfnamefont {Dafei}\ \bibnamefont {Jin}}, \bibinfo {author} {\bibfnamefont {Yang}\ \bibnamefont {Ma}}, \bibinfo {author} {\bibfnamefont {Haw-Wen}\ \bibnamefont {Hsiao}}, \bibinfo {author} {\bibfnamefont {Yulin}\ \bibnamefont {Lin}}, \bibinfo {author} {\bibfnamefont {Terence~M.}\ \bibnamefont {Bretz-Sullivan}}, \bibinfo {author} {\bibfnamefont {Xianjing}\ \bibnamefont {Zhou}}, \bibinfo {author} {\bibfnamefont {John}\ \bibnamefont {Pearson}}, \bibinfo {author} {\bibfnamefont {Brandon}\ \bibnamefont {Fisher}}, \bibinfo {author} {\bibfnamefont {J.~Samuel}\ \bibnamefont {Jiang}}, \bibinfo {author} {\bibfnamefont {Wei}\ \bibnamefont {Han}}, \bibinfo {author} {\bibfnamefont {Jian-Min}\ \bibnamefont {Zuo}}, \bibinfo {author} {\bibfnamefont {Jianguo}\ \bibnamefont {Wen}}, \bibinfo {author} {\bibfnamefont {Dillon~D.}\ \bibnamefont {Fong}}, \bibinfo
  {author} {\bibfnamefont {Jirong}\ \bibnamefont {Sun}}, \bibinfo {author} {\bibfnamefont {Hua}\ \bibnamefont {Zhou}}, \ and\ \bibinfo {author} {\bibfnamefont {Anand}\ \bibnamefont {Bhattacharya}},\ }\bibfield  {title} {\enquote {\bibinfo {title} {Two-dimensional superconductivity and anisotropic transport at $\mathrm{KTaO_3}$ (111) interfaces},}\ }\href {\doibase 10.1126/science.aba5511} {\bibfield  {journal} {\bibinfo  {journal} {Science}\ }\textbf {\bibinfo {volume} {371}},\ \bibinfo {pages} {716--721} (\bibinfo {year} {2021})}\BibitemShut {NoStop}%
\bibitem [{\citenamefont {Chen}\ \emph {et~al.}(2021)\citenamefont {Chen}, \citenamefont {Liu}, \citenamefont {Zhang}, \citenamefont {Liu}, \citenamefont {Tian}, \citenamefont {Sun}, \citenamefont {Zhang}, \citenamefont {Zhou}, \citenamefont {Sun},\ and\ \citenamefont {Xie}}]{kto2}%
  \BibitemOpen
  \bibfield  {author} {\bibinfo {author} {\bibfnamefont {Zheng}\ \bibnamefont {Chen}}, \bibinfo {author} {\bibfnamefont {Yuan}\ \bibnamefont {Liu}}, \bibinfo {author} {\bibfnamefont {Hui}\ \bibnamefont {Zhang}}, \bibinfo {author} {\bibfnamefont {Zhongran}\ \bibnamefont {Liu}}, \bibinfo {author} {\bibfnamefont {He}~\bibnamefont {Tian}}, \bibinfo {author} {\bibfnamefont {Yanqiu}\ \bibnamefont {Sun}}, \bibinfo {author} {\bibfnamefont {Meng}\ \bibnamefont {Zhang}}, \bibinfo {author} {\bibfnamefont {Yi}~\bibnamefont {Zhou}}, \bibinfo {author} {\bibfnamefont {Jirong}\ \bibnamefont {Sun}}, \ and\ \bibinfo {author} {\bibfnamefont {Yanwu}\ \bibnamefont {Xie}},\ }\bibfield  {title} {\enquote {\bibinfo {title} {Electric field control of superconductivity at the $\mathrm{LaAlO_3/KTaO_3}$(111) interface},}\ }\href {\doibase 10.1126/science.abb3848} {\bibfield  {journal} {\bibinfo  {journal} {Science}\ }\textbf {\bibinfo {volume} {372}},\ \bibinfo {pages} {721--724} (\bibinfo {year} {2021})}\BibitemShut {NoStop}%
\bibitem [{\citenamefont {Zhang}\ \emph {et~al.}(2018)\citenamefont {Zhang}, \citenamefont {Yun}, \citenamefont {Zhang}, \citenamefont {Zhang}, \citenamefont {Ma}, \citenamefont {Yan}, \citenamefont {Wang}, \citenamefont {Li}, \citenamefont {Li}, \citenamefont {Khan}, \citenamefont {Chen}, \citenamefont {Liu}, \citenamefont {Hu}, \citenamefont {Liu}, \citenamefont {Shen}, \citenamefont {Han},\ and\ \citenamefont {Sun}}]{PhysRevLett.121.116803}%
  \BibitemOpen
  \bibfield  {author} {\bibinfo {author} {\bibfnamefont {Hongrui}\ \bibnamefont {Zhang}}, \bibinfo {author} {\bibfnamefont {Yu}~\bibnamefont {Yun}}, \bibinfo {author} {\bibfnamefont {Xuejing}\ \bibnamefont {Zhang}}, \bibinfo {author} {\bibfnamefont {Hui}\ \bibnamefont {Zhang}}, \bibinfo {author} {\bibfnamefont {Yang}\ \bibnamefont {Ma}}, \bibinfo {author} {\bibfnamefont {Xi}~\bibnamefont {Yan}}, \bibinfo {author} {\bibfnamefont {Fei}\ \bibnamefont {Wang}}, \bibinfo {author} {\bibfnamefont {Gang}\ \bibnamefont {Li}}, \bibinfo {author} {\bibfnamefont {Rui}\ \bibnamefont {Li}}, \bibinfo {author} {\bibfnamefont {Tahira}\ \bibnamefont {Khan}}, \bibinfo {author} {\bibfnamefont {Yuansha}\ \bibnamefont {Chen}}, \bibinfo {author} {\bibfnamefont {Wei}\ \bibnamefont {Liu}}, \bibinfo {author} {\bibfnamefont {Fengxia}\ \bibnamefont {Hu}}, \bibinfo {author} {\bibfnamefont {Banggui}\ \bibnamefont {Liu}}, \bibinfo {author} {\bibfnamefont {Baogen}\ \bibnamefont {Shen}}, \bibinfo {author} {\bibfnamefont {Wei}\ \bibnamefont
  {Han}}, \ and\ \bibinfo {author} {\bibfnamefont {Jirong}\ \bibnamefont {Sun}},\ }\bibfield  {title} {\enquote {\bibinfo {title} {High-mobility spin-polarized two-dimensional electron gases at $\mathrm{EuO}/\mathrm{KTaO}_{3}$ interfaces},}\ }\href {\doibase 10.1103/PhysRevLett.121.116803} {\bibfield  {journal} {\bibinfo  {journal} {Phys. Rev. Lett.}\ }\textbf {\bibinfo {volume} {121}},\ \bibinfo {pages} {116803} (\bibinfo {year} {2018})}\BibitemShut {NoStop}%
\bibitem [{\citenamefont {Qiao}\ \emph {et~al.}(2021)\citenamefont {Qiao}, \citenamefont {Ma}, \citenamefont {Yan}, \citenamefont {Xing}, \citenamefont {Yao}, \citenamefont {Cai}, \citenamefont {Li}, \citenamefont {Xiong}, \citenamefont {Xie}, \citenamefont {Lin},\ and\ \citenamefont {Han}}]{PhysRevB.104.184505}%
  \BibitemOpen
  \bibfield  {author} {\bibinfo {author} {\bibfnamefont {Weiliang}\ \bibnamefont {Qiao}}, \bibinfo {author} {\bibfnamefont {Yang}\ \bibnamefont {Ma}}, \bibinfo {author} {\bibfnamefont {Jiaojie}\ \bibnamefont {Yan}}, \bibinfo {author} {\bibfnamefont {Wenyu}\ \bibnamefont {Xing}}, \bibinfo {author} {\bibfnamefont {Yunyan}\ \bibnamefont {Yao}}, \bibinfo {author} {\bibfnamefont {Ranran}\ \bibnamefont {Cai}}, \bibinfo {author} {\bibfnamefont {Boning}\ \bibnamefont {Li}}, \bibinfo {author} {\bibfnamefont {Richen}\ \bibnamefont {Xiong}}, \bibinfo {author} {\bibfnamefont {X.~C.}\ \bibnamefont {Xie}}, \bibinfo {author} {\bibfnamefont {Xi}~\bibnamefont {Lin}}, \ and\ \bibinfo {author} {\bibfnamefont {Wei}\ \bibnamefont {Han}},\ }\bibfield  {title} {\enquote {\bibinfo {title} {Gate tunability of the superconducting state at the $\mathrm{EuO}/\mathrm{KTa}\mathrm{O}_{3}$ (111) interface},}\ }\href {\doibase 10.1103/PhysRevB.104.184505} {\bibfield  {journal} {\bibinfo  {journal} {Phys. Rev. B}\ }\textbf {\bibinfo {volume}
  {104}},\ \bibinfo {pages} {184505} (\bibinfo {year} {2021})}\BibitemShut {NoStop}%
\bibitem [{\citenamefont {Hua}\ \emph {et~al.}(2022)\citenamefont {Hua}, \citenamefont {Meng}, \citenamefont {Huang}, \citenamefont {Li}, \citenamefont {Wang}, \citenamefont {Ge}, \citenamefont {Xiang},\ and\ \citenamefont {Chen}}]{Hua2022}%
  \BibitemOpen
  \bibfield  {author} {\bibinfo {author} {\bibfnamefont {Xiangyu}\ \bibnamefont {Hua}}, \bibinfo {author} {\bibfnamefont {Fanbao}\ \bibnamefont {Meng}}, \bibinfo {author} {\bibfnamefont {Zongyao}\ \bibnamefont {Huang}}, \bibinfo {author} {\bibfnamefont {Zhaohang}\ \bibnamefont {Li}}, \bibinfo {author} {\bibfnamefont {Shuai}\ \bibnamefont {Wang}}, \bibinfo {author} {\bibfnamefont {Binghui}\ \bibnamefont {Ge}}, \bibinfo {author} {\bibfnamefont {Ziji}\ \bibnamefont {Xiang}}, \ and\ \bibinfo {author} {\bibfnamefont {Xianhui}\ \bibnamefont {Chen}},\ }\bibfield  {title} {\enquote {\bibinfo {title} {Tunable two-dimensional superconductivity and spin-orbit coupling at the $\mathrm{EuO/KTaO_3}$(110) interface},}\ }\href {\doibase 10.1038/s41535-022-00506-x} {\bibfield  {journal} {\bibinfo  {journal} {npj Quantum Materials}\ }\textbf {\bibinfo {volume} {7}},\ \bibinfo {pages} {97} (\bibinfo {year} {2022})}\BibitemShut {NoStop}%
\bibitem [{\citenamefont {Liu}\ \emph {et~al.}(2023)\citenamefont {Liu}, \citenamefont {Zhou}, \citenamefont {Hong}, \citenamefont {Fisher}, \citenamefont {Zheng}, \citenamefont {Pearson}, \citenamefont {Jiang}, \citenamefont {Jin}, \citenamefont {Norman},\ and\ \citenamefont {Bhattacharya}}]{Liu2023}%
  \BibitemOpen
  \bibfield  {author} {\bibinfo {author} {\bibfnamefont {Changjiang}\ \bibnamefont {Liu}}, \bibinfo {author} {\bibfnamefont {Xianjing}\ \bibnamefont {Zhou}}, \bibinfo {author} {\bibfnamefont {Deshun}\ \bibnamefont {Hong}}, \bibinfo {author} {\bibfnamefont {Brandon}\ \bibnamefont {Fisher}}, \bibinfo {author} {\bibfnamefont {Hong}\ \bibnamefont {Zheng}}, \bibinfo {author} {\bibfnamefont {John}\ \bibnamefont {Pearson}}, \bibinfo {author} {\bibfnamefont {Jidong~Samuel}\ \bibnamefont {Jiang}}, \bibinfo {author} {\bibfnamefont {Dafei}\ \bibnamefont {Jin}}, \bibinfo {author} {\bibfnamefont {Michael~R.}\ \bibnamefont {Norman}}, \ and\ \bibinfo {author} {\bibfnamefont {Anand}\ \bibnamefont {Bhattacharya}},\ }\bibfield  {title} {\enquote {\bibinfo {title} {Tunable superconductivity and its origin at $\mathrm{KTaO_3}$ interfaces},}\ }\href {\doibase 10.1038/s41467-023-36309-2} {\bibfield  {journal} {\bibinfo  {journal} {Nature Communications}\ }\textbf {\bibinfo {volume} {14}},\ \bibinfo {pages} {951} (\bibinfo {year}
  {2023})}\BibitemShut {NoStop}%
\bibitem [{\citenamefont {Hua}\ \emph {et~al.}(2024)\citenamefont {Hua}, \citenamefont {Zeng}, \citenamefont {Meng}, \citenamefont {Yao}, \citenamefont {Huang}, \citenamefont {Long}, \citenamefont {Li}, \citenamefont {Wang}, \citenamefont {Wang}, \citenamefont {Wu}, \citenamefont {Weng}, \citenamefont {Wang}, \citenamefont {Liu}, \citenamefont {Xiang},\ and\ \citenamefont {Chen}}]{Hua2024}%
  \BibitemOpen
  \bibfield  {author} {\bibinfo {author} {\bibfnamefont {Xiangyu}\ \bibnamefont {Hua}}, \bibinfo {author} {\bibfnamefont {Zimeng}\ \bibnamefont {Zeng}}, \bibinfo {author} {\bibfnamefont {Fanbao}\ \bibnamefont {Meng}}, \bibinfo {author} {\bibfnamefont {Hongxu}\ \bibnamefont {Yao}}, \bibinfo {author} {\bibfnamefont {Zongyao}\ \bibnamefont {Huang}}, \bibinfo {author} {\bibfnamefont {Xuanyu}\ \bibnamefont {Long}}, \bibinfo {author} {\bibfnamefont {Zhaohang}\ \bibnamefont {Li}}, \bibinfo {author} {\bibfnamefont {Youfang}\ \bibnamefont {Wang}}, \bibinfo {author} {\bibfnamefont {Zhenyu}\ \bibnamefont {Wang}}, \bibinfo {author} {\bibfnamefont {Tao}\ \bibnamefont {Wu}}, \bibinfo {author} {\bibfnamefont {Zhengyu}\ \bibnamefont {Weng}}, \bibinfo {author} {\bibfnamefont {Yihua}\ \bibnamefont {Wang}}, \bibinfo {author} {\bibfnamefont {Zheng}\ \bibnamefont {Liu}}, \bibinfo {author} {\bibfnamefont {Ziji}\ \bibnamefont {Xiang}}, \ and\ \bibinfo {author} {\bibfnamefont {Xianhui}\ \bibnamefont {Chen}},\ }\bibfield  {title}
  {\enquote {\bibinfo {title} {Superconducting stripes induced by ferromagnetic proximity in an oxide heterostructure},}\ }\href {\doibase 10.1038/s41567-024-02443-x} {\bibfield  {journal} {\bibinfo  {journal} {Nature Physics}\ }\textbf {\bibinfo {volume} {20}},\ \bibinfo {pages} {957--963} (\bibinfo {year} {2024})}\BibitemShut {NoStop}%
\bibitem [{\citenamefont {Cao}\ \emph {et~al.}(2018)\citenamefont {Cao}, \citenamefont {Fatemi}, \citenamefont {Fang}, \citenamefont {Watanabe}, \citenamefont {Taniguchi}, \citenamefont {Kaxiras},\ and\ \citenamefont {Jarillo-Herrero}}]{bilayer-graphene}%
  \BibitemOpen
  \bibfield  {author} {\bibinfo {author} {\bibfnamefont {Yuan}\ \bibnamefont {Cao}}, \bibinfo {author} {\bibfnamefont {Valla}\ \bibnamefont {Fatemi}}, \bibinfo {author} {\bibfnamefont {Shiang}\ \bibnamefont {Fang}}, \bibinfo {author} {\bibfnamefont {Kenji}\ \bibnamefont {Watanabe}}, \bibinfo {author} {\bibfnamefont {Takashi}\ \bibnamefont {Taniguchi}}, \bibinfo {author} {\bibfnamefont {Efthimios}\ \bibnamefont {Kaxiras}}, \ and\ \bibinfo {author} {\bibfnamefont {Pablo}\ \bibnamefont {Jarillo-Herrero}},\ }\bibfield  {title} {\enquote {\bibinfo {title} {Unconventional superconductivity in magic-angle graphene superlattices},}\ }\href {\doibase 10.1038/nature26160} {\bibfield  {journal} {\bibinfo  {journal} {Nature}\ }\textbf {\bibinfo {volume} {556}},\ \bibinfo {pages} {43--50} (\bibinfo {year} {2018})}\BibitemShut {NoStop}%
\bibitem [{\citenamefont {Saito}\ \emph {et~al.}(2016)\citenamefont {Saito}, \citenamefont {Nojima},\ and\ \citenamefont {Iwasa}}]{2d-SC-review}%
  \BibitemOpen
  \bibfield  {author} {\bibinfo {author} {\bibfnamefont {Yu}~\bibnamefont {Saito}}, \bibinfo {author} {\bibfnamefont {Tsutomu}\ \bibnamefont {Nojima}}, \ and\ \bibinfo {author} {\bibfnamefont {Yoshihiro}\ \bibnamefont {Iwasa}},\ }\bibfield  {title} {\enquote {\bibinfo {title} {Highly crystalline 2d superconductors},}\ }\href {\doibase 10.1038/natrevmats.2016.94} {\bibfield  {journal} {\bibinfo  {journal} {Nature Reviews Materials}\ }\textbf {\bibinfo {volume} {2}},\ \bibinfo {pages} {16094} (\bibinfo {year} {2016})}\BibitemShut {NoStop}%
\bibitem [{\citenamefont {Berezinsky}(1971)}]{Berezinsky:1970fr}%
  \BibitemOpen
  \bibfield  {author} {\bibinfo {author} {\bibfnamefont {V.~L.}\ \bibnamefont {Berezinskii}},\ }\bibfield  {title} {\enquote {\bibinfo {title} {{Destruction of long-range order in one-dimensional and two-dimensional systems having a continuous symmetry group I. Classical systems}},}\ }\href@noop {} {\bibfield  {journal} {\bibinfo  {journal} {Sov. Phys. JETP}\ }\textbf {\bibinfo {volume} {32}},\ \bibinfo {pages} {493--500} (\bibinfo {year} {1971})}\BibitemShut {NoStop}%
\bibitem [{\citenamefont {Kosterlitz}\ and\ \citenamefont {Thouless}(1973)}]{Kosterlitz_1973}%
  \BibitemOpen
  \bibfield  {author} {\bibinfo {author} {\bibfnamefont {J~M}\ \bibnamefont {Kosterlitz}}\ and\ \bibinfo {author} {\bibfnamefont {D~J}\ \bibnamefont {Thouless}},\ }\bibfield  {title} {\enquote {\bibinfo {title} {Ordering, metastability and phase transitions in two-dimensional systems},}\ }\href {\doibase 10.1088/0022-3719/6/7/010} {\bibfield  {journal} {\bibinfo  {journal} {Journal of Physics C: Solid State Physics}\ }\textbf {\bibinfo {volume} {6}},\ \bibinfo {pages} {1181} (\bibinfo {year} {1973})}\BibitemShut {NoStop}%
\bibitem [{\citenamefont {Kosterlitz}(1974)}]{Kosterlitz_1974}%
  \BibitemOpen
  \bibfield  {author} {\bibinfo {author} {\bibfnamefont {J~M}\ \bibnamefont {Kosterlitz}},\ }\bibfield  {title} {\enquote {\bibinfo {title} {The critical properties of the two-dimensional xy model},}\ }\href {\doibase 10.1088/0022-3719/7/6/005} {\bibfield  {journal} {\bibinfo  {journal} {Journal of Physics C: Solid State Physics}\ }\textbf {\bibinfo {volume} {7}},\ \bibinfo {pages} {1046} (\bibinfo {year} {1974})}\BibitemShut {NoStop}%
\bibitem [{\citenamefont {Halperin}\ and\ \citenamefont {Nelson}(1979)}]{Halperin1979}%
  \BibitemOpen
  \bibfield  {author} {\bibinfo {author} {\bibfnamefont {B.~I.}\ \bibnamefont {Halperin}}\ and\ \bibinfo {author} {\bibfnamefont {David~R.}\ \bibnamefont {Nelson}},\ }\bibfield  {title} {\enquote {\bibinfo {title} {Resistive transition in superconducting films},}\ }\href {\doibase 10.1007/BF00116988} {\bibfield  {journal} {\bibinfo  {journal} {Journal of Low Temperature Physics}\ }\textbf {\bibinfo {volume} {36}},\ \bibinfo {pages} {599--616} (\bibinfo {year} {1979})}\BibitemShut {NoStop}%
\bibitem [{\citenamefont {Li}\ \emph {et~al.}(2024)\citenamefont {Li}, \citenamefont {Kivelson},\ and\ \citenamefont {Lee}}]{li2024theoryinfinitelyanisotropicphase}%
  \BibitemOpen
  \bibfield  {author} {\bibinfo {author} {\bibfnamefont {Zi-Xiang}\ \bibnamefont {Li}}, \bibinfo {author} {\bibfnamefont {Steven~A}\ \bibnamefont {Kivelson}}, \ and\ \bibinfo {author} {\bibfnamefont {Dung-Hai}\ \bibnamefont {Lee}},\ }\bibfield  {title} {\enquote {\bibinfo {title} {Theory of an infinitely anisotropic phase of a two-dimensional superconductor},}\ }\href {https://arxiv.org/abs/2407.10269} {\bibfield  {journal} {\bibinfo  {journal} {arXiv:2407.10269}\ } (\bibinfo {year} {2024})}\BibitemShut {NoStop}%
\bibitem [{\citenamefont {Tinkham}(1996)}]{tinkham1996introduction}%
  \BibitemOpen
  \bibfield  {author} {\bibinfo {author} {\bibfnamefont {M.}~\bibnamefont {Tinkham}},\ }\href {https://books.google.com.hk/books?id=XP_uAAAAMAAJ} {\emph {\bibinfo {title} {Introduction to Superconductivity}}}\ (\bibinfo  {publisher} {McGraw Hill},\ \bibinfo {year} {1996})\BibitemShut {NoStop}%
\bibitem [{\citenamefont {Kopnin}(2001)}]{kopnin2001theory}%
  \BibitemOpen
  \bibfield  {author} {\bibinfo {author} {\bibfnamefont {N.~B.}\ \bibnamefont {Kopnin}},\ }\href {https://books.google.com.tw/books?id=VuA-DgAAQBAJ} {\emph {\bibinfo {title} {Theory of Nonequilibrium Superconductivity}}}\ (\bibinfo  {publisher} {Clarendon Press},\ \bibinfo {year} {2001})\BibitemShut {NoStop}%
\bibitem [{\citenamefont {Bardeen}\ and\ \citenamefont {Stephen}(1965)}]{bardeen-stephen}%
  \BibitemOpen
  \bibfield  {author} {\bibinfo {author} {\bibfnamefont {John}\ \bibnamefont {Bardeen}}\ and\ \bibinfo {author} {\bibfnamefont {M.~J.}\ \bibnamefont {Stephen}},\ }\bibfield  {title} {\enquote {\bibinfo {title} {Theory of the motion of vortices in superconductors},}\ }\href {\doibase 10.1103/PhysRev.140.A1197} {\bibfield  {journal} {\bibinfo  {journal} {Phys. Rev.}\ }\textbf {\bibinfo {volume} {140}},\ \bibinfo {pages} {A1197--A1207} (\bibinfo {year} {1965})}\BibitemShut {NoStop}%
\bibitem [{\citenamefont {Tinkham}(1964)}]{Tinkham_PhysRevLett.13.804}%
  \BibitemOpen
  \bibfield  {author} {\bibinfo {author} {\bibfnamefont {M.}~\bibnamefont {Tinkham}},\ }\bibfield  {title} {\enquote {\bibinfo {title} {Viscous flow of flux in type-ii superconductors},}\ }\href {\doibase 10.1103/PhysRevLett.13.804} {\bibfield  {journal} {\bibinfo  {journal} {Phys. Rev. Lett.}\ }\textbf {\bibinfo {volume} {13}},\ \bibinfo {pages} {804--807} (\bibinfo {year} {1964})}\BibitemShut {NoStop}%
\bibitem [{\citenamefont {Carlson}\ \emph {et~al.}(2003)\citenamefont {Carlson}, \citenamefont {Castro~Neto},\ and\ \citenamefont {Campbell}}]{PhysRevLett.90.087001}%
  \BibitemOpen
  \bibfield  {author} {\bibinfo {author} {\bibfnamefont {E.~W.}\ \bibnamefont {Carlson}}, \bibinfo {author} {\bibfnamefont {A.~H.}\ \bibnamefont {Castro~Neto}}, \ and\ \bibinfo {author} {\bibfnamefont {D.~K.}\ \bibnamefont {Campbell}},\ }\bibfield  {title} {\enquote {\bibinfo {title} {Vortex liquid crystals in anisotropic type {II} superconductors},}\ }\href {\doibase 10.1103/PhysRevLett.90.087001} {\bibfield  {journal} {\bibinfo  {journal} {Phys. Rev. Lett.}\ }\textbf {\bibinfo {volume} {90}},\ \bibinfo {pages} {087001} (\bibinfo {year} {2003})}\BibitemShut {NoStop}%
\bibitem [{\citenamefont {Reichhardt}\ and\ \citenamefont {Reichhardt}(2006)}]{C.Reichhardt_2006}%
  \BibitemOpen
  \bibfield  {author} {\bibinfo {author} {\bibfnamefont {C.}~\bibnamefont {Reichhardt}}\ and\ \bibinfo {author} {\bibfnamefont {C.~J.~Olson}\ \bibnamefont {Reichhardt}},\ }\bibfield  {title} {\enquote {\bibinfo {title} {Statics and dynamics of two-dimensional vortex liquid crystals},}\ }\href {\doibase 10.1209/epl/i2005-10602-4} {\bibfield  {journal} {\bibinfo  {journal} {Europhysics Letters}\ }\textbf {\bibinfo {volume} {75}},\ \bibinfo {pages} {489} (\bibinfo {year} {2006})}\BibitemShut {NoStop}%
\bibitem [{\citenamefont {Guillamón}\ \emph {et~al.}(2009)\citenamefont {Guillamón}, \citenamefont {Suderow}, \citenamefont {Fernández-Pacheco}, \citenamefont {Sesé}, \citenamefont {Córdoba}, \citenamefont {De~Teresa}, \citenamefont {Ibarra},\ and\ \citenamefont {Vieira}}]{Guillamon2009}%
  \BibitemOpen
  \bibfield  {author} {\bibinfo {author} {\bibfnamefont {I.}~\bibnamefont {Guillamón}}, \bibinfo {author} {\bibfnamefont {H.}~\bibnamefont {Suderow}}, \bibinfo {author} {\bibfnamefont {A.}~\bibnamefont {Fernández-Pacheco}}, \bibinfo {author} {\bibfnamefont {J.}~\bibnamefont {Sesé}}, \bibinfo {author} {\bibfnamefont {R.}~\bibnamefont {Córdoba}}, \bibinfo {author} {\bibfnamefont {J.~M.}\ \bibnamefont {De~Teresa}}, \bibinfo {author} {\bibfnamefont {M.~R.}\ \bibnamefont {Ibarra}}, \ and\ \bibinfo {author} {\bibfnamefont {S.}~\bibnamefont {Vieira}},\ }\bibfield  {title} {\enquote {\bibinfo {title} {Direct observation of melting in a two-dimensional superconducting vortex lattice},}\ }\href {\doibase 10.1038/nphys1368} {\bibfield  {journal} {\bibinfo  {journal} {Nature Physics}\ }\textbf {\bibinfo {volume} {5}},\ \bibinfo {pages} {651--655} (\bibinfo {year} {2009})}\BibitemShut {NoStop}%
\bibitem [{\citenamefont {Guillamón}\ \emph {et~al.}(2014)\citenamefont {Guillamón}, \citenamefont {Córdoba}, \citenamefont {Sesé}, \citenamefont {De~Teresa}, \citenamefont {Ibarra}, \citenamefont {Vieira},\ and\ \citenamefont {Suderow}}]{Guillamon2014}%
  \BibitemOpen
  \bibfield  {author} {\bibinfo {author} {\bibfnamefont {I.}~\bibnamefont {Guillamón}}, \bibinfo {author} {\bibfnamefont {R.}~\bibnamefont {Córdoba}}, \bibinfo {author} {\bibfnamefont {J.}~\bibnamefont {Sesé}}, \bibinfo {author} {\bibfnamefont {J.~M.}\ \bibnamefont {De~Teresa}}, \bibinfo {author} {\bibfnamefont {M.~R.}\ \bibnamefont {Ibarra}}, \bibinfo {author} {\bibfnamefont {S.}~\bibnamefont {Vieira}}, \ and\ \bibinfo {author} {\bibfnamefont {H.}~\bibnamefont {Suderow}},\ }\bibfield  {title} {\enquote {\bibinfo {title} {Enhancement of long-range correlations in a {2D} vortex lattice by an incommensurate 1d disorder potential},}\ }\href {\doibase 10.1038/nphys3132} {\bibfield  {journal} {\bibinfo  {journal} {Nature Physics}\ }\textbf {\bibinfo {volume} {10}},\ \bibinfo {pages} {851--856} (\bibinfo {year} {2014})}\BibitemShut {NoStop}%
\bibitem [{\citenamefont {Le~Thien}\ \emph {et~al.}(2016)\citenamefont {Le~Thien}, \citenamefont {McDermott}, \citenamefont {Olson~Reichhardt},\ and\ \citenamefont {Reichhardt}}]{PhysRevB.93.014504}%
  \BibitemOpen
  \bibfield  {author} {\bibinfo {author} {\bibfnamefont {Q.}~\bibnamefont {Le~Thien}}, \bibinfo {author} {\bibfnamefont {D.}~\bibnamefont {McDermott}}, \bibinfo {author} {\bibfnamefont {C.~J.}\ \bibnamefont {Olson~Reichhardt}}, \ and\ \bibinfo {author} {\bibfnamefont {C.}~\bibnamefont {Reichhardt}},\ }\bibfield  {title} {\enquote {\bibinfo {title} {Orientational ordering, buckling, and dynamic transitions for vortices interacting with a periodic quasi-one-dimensional substrate},}\ }\href {\doibase 10.1103/PhysRevB.93.014504} {\bibfield  {journal} {\bibinfo  {journal} {Phys. Rev. B}\ }\textbf {\bibinfo {volume} {93}},\ \bibinfo {pages} {014504} (\bibinfo {year} {2016})}\BibitemShut {NoStop}%
\bibitem [{\citenamefont {Dobrovolskiy}\ \emph {et~al.}(2020)\citenamefont {Dobrovolskiy}, \citenamefont {Vodolazov}, \citenamefont {Porrati}, \citenamefont {Sachser}, \citenamefont {Bevz}, \citenamefont {Mikhailov}, \citenamefont {Chumak},\ and\ \citenamefont {Huth}}]{Dobrovolskiy2020}%
  \BibitemOpen
  \bibfield  {author} {\bibinfo {author} {\bibfnamefont {O.~V.}\ \bibnamefont {Dobrovolskiy}}, \bibinfo {author} {\bibfnamefont {D.~Yu}\ \bibnamefont {Vodolazov}}, \bibinfo {author} {\bibfnamefont {F.}~\bibnamefont {Porrati}}, \bibinfo {author} {\bibfnamefont {R.}~\bibnamefont {Sachser}}, \bibinfo {author} {\bibfnamefont {V.~M.}\ \bibnamefont {Bevz}}, \bibinfo {author} {\bibfnamefont {M.~Yu}\ \bibnamefont {Mikhailov}}, \bibinfo {author} {\bibfnamefont {A.~V.}\ \bibnamefont {Chumak}}, \ and\ \bibinfo {author} {\bibfnamefont {M.}~\bibnamefont {Huth}},\ }\bibfield  {title} {\enquote {\bibinfo {title} {Ultra-fast vortex motion in a direct-write {Nb-C} superconductor},}\ }\href {\doibase 10.1038/s41467-020-16987-y} {\bibfield  {journal} {\bibinfo  {journal} {Nature Communications}\ }\textbf {\bibinfo {volume} {11}},\ \bibinfo {pages} {3291} (\bibinfo {year} {2020})}\BibitemShut {NoStop}%
\bibitem [{\citenamefont {Berezinskii}(1972)}]{Berezinskii:1972fet}%
  \BibitemOpen
  \bibfield  {author} {\bibinfo {author} {\bibfnamefont {V.~L.}\ \bibnamefont {Berezinskii}},\ }\bibfield  {title} {\enquote {\bibinfo {title} {{Destruction of Long-range Order in One-dimensional and Two-dimensional Systems Possessing a Continuous Symmetry Group. II. Quantum Systems}},}\ }\href@noop {} {\bibfield  {journal} {\bibinfo  {journal} {Sov. Phys. JETP}\ }\textbf {\bibinfo {volume} {34}},\ \bibinfo {pages} {610--616} (\bibinfo {year} {1972})}\BibitemShut {NoStop}%
\bibitem [{\citenamefont {Halperin}\ and\ \citenamefont {Nelson}(1978)}]{PhysRevLett.41.121}%
  \BibitemOpen
  \bibfield  {author} {\bibinfo {author} {\bibfnamefont {B.~I.}\ \bibnamefont {Halperin}}\ and\ \bibinfo {author} {\bibfnamefont {David~R.}\ \bibnamefont {Nelson}},\ }\bibfield  {title} {\enquote {\bibinfo {title} {Theory of two-dimensional melting},}\ }\href {\doibase 10.1103/PhysRevLett.41.121} {\bibfield  {journal} {\bibinfo  {journal} {Phys. Rev. Lett.}\ }\textbf {\bibinfo {volume} {41}},\ \bibinfo {pages} {121--124} (\bibinfo {year} {1978})}\BibitemShut {NoStop}%
\bibitem [{\citenamefont {Nelson}\ and\ \citenamefont {Halperin}(1979)}]{PhysRevB.19.2457}%
  \BibitemOpen
  \bibfield  {author} {\bibinfo {author} {\bibfnamefont {David~R.}\ \bibnamefont {Nelson}}\ and\ \bibinfo {author} {\bibfnamefont {B.~I.}\ \bibnamefont {Halperin}},\ }\bibfield  {title} {\enquote {\bibinfo {title} {Dislocation-mediated melting in two dimensions},}\ }\href {\doibase 10.1103/PhysRevB.19.2457} {\bibfield  {journal} {\bibinfo  {journal} {Phys. Rev. B}\ }\textbf {\bibinfo {volume} {19}},\ \bibinfo {pages} {2457--2484} (\bibinfo {year} {1979})}\BibitemShut {NoStop}%
\bibitem [{\citenamefont {Hoshino}\ \emph {et~al.}(2018)\citenamefont {Hoshino}, \citenamefont {Wakatsuki}, \citenamefont {Hamamoto},\ and\ \citenamefont {Nagaosa}}]{PhysRevB.98.054510}%
  \BibitemOpen
  \bibfield  {author} {\bibinfo {author} {\bibfnamefont {Shintaro}\ \bibnamefont {Hoshino}}, \bibinfo {author} {\bibfnamefont {Ryohei}\ \bibnamefont {Wakatsuki}}, \bibinfo {author} {\bibfnamefont {Keita}\ \bibnamefont {Hamamoto}}, \ and\ \bibinfo {author} {\bibfnamefont {Naoto}\ \bibnamefont {Nagaosa}},\ }\bibfield  {title} {\enquote {\bibinfo {title} {Nonreciprocal charge transport in two-dimensional noncentrosymmetric superconductors},}\ }\href {\doibase 10.1103/PhysRevB.98.054510} {\bibfield  {journal} {\bibinfo  {journal} {Phys. Rev. B}\ }\textbf {\bibinfo {volume} {98}},\ \bibinfo {pages} {054510} (\bibinfo {year} {2018})}\BibitemShut {NoStop}%
\bibitem [{\citenamefont {Zwanzig}(2001)}]{zwanzig2001nonequilibrium}%
  \BibitemOpen
  \bibfield  {author} {\bibinfo {author} {\bibfnamefont {R.}~\bibnamefont {Zwanzig}},\ }\href {https://books.google.com.hk/books?id=4cI5136OdoMC} {\emph {\bibinfo {title} {Nonequilibrium Statistical Mechanics}}}\ (\bibinfo  {publisher} {Oxford University Press},\ \bibinfo {year} {2001})\BibitemShut {NoStop}%
\bibitem [{Sup()}]{Supplement}%
  \BibitemOpen
  \href@noop {} {}\bibinfo {note} {See Supplemental Material at [http://link.aps.org/supplemental/
  10.1103/vb55-n7rr] for the details of the 2D simulation for the BKT case, an analytical treatment of the probing-current dependence of the BKT transition anisotropy, the numerical simulations of the probing-current dependence of the transition anisotropy, and the argument for the neglect of transverse force terms in the vortex dynamics.}\BibitemShut {Stop}%
\bibitem [{\citenamefont {Koshelev}\ and\ \citenamefont {Vinokur}(1994)}]{PhysRevLett.73.3580}%
  \BibitemOpen
  \bibfield  {author} {\bibinfo {author} {\bibfnamefont {A.~E.}\ \bibnamefont {Koshelev}}\ and\ \bibinfo {author} {\bibfnamefont {V.~M.}\ \bibnamefont {Vinokur}},\ }\bibfield  {title} {\enquote {\bibinfo {title} {Dynamic melting of the vortex lattice},}\ }\href {\doibase 10.1103/PhysRevLett.73.3580} {\bibfield  {journal} {\bibinfo  {journal} {Phys. Rev. Lett.}\ }\textbf {\bibinfo {volume} {73}},\ \bibinfo {pages} {3580--3583} (\bibinfo {year} {1994})}\BibitemShut {NoStop}%
\bibitem [{\citenamefont {Erta\ifmmode~\mbox{\c{s}}\else \c{s}\fi{}}\ and\ \citenamefont {Kardar}(1996)}]{PhysRevB.53.3520}%
  \BibitemOpen
  \bibfield  {author} {\bibinfo {author} {\bibfnamefont {Deniz}\ \bibnamefont {Erta\ifmmode~\mbox{\c{s}}\else \c{s}\fi{}}}\ and\ \bibinfo {author} {\bibfnamefont {Mehran}\ \bibnamefont {Kardar}},\ }\bibfield  {title} {\enquote {\bibinfo {title} {Anisotropic scaling in threshold critical dynamics of driven directed lines},}\ }\href {\doibase 10.1103/PhysRevB.53.3520} {\bibfield  {journal} {\bibinfo  {journal} {Phys. Rev. B}\ }\textbf {\bibinfo {volume} {53}},\ \bibinfo {pages} {3520--3542} (\bibinfo {year} {1996})}\BibitemShut {NoStop}%
\bibitem [{\citenamefont {Bustingorry}\ \emph {et~al.}(2007)\citenamefont {Bustingorry}, \citenamefont {Cugliandolo},\ and\ \citenamefont {Dom\'{\i}nguez}}]{PhysRevB.75.024506}%
  \BibitemOpen
  \bibfield  {author} {\bibinfo {author} {\bibfnamefont {Sebastian}\ \bibnamefont {Bustingorry}}, \bibinfo {author} {\bibfnamefont {Leticia~F.}\ \bibnamefont {Cugliandolo}}, \ and\ \bibinfo {author} {\bibfnamefont {Daniel}\ \bibnamefont {Dom\'{\i}nguez}},\ }\bibfield  {title} {\enquote {\bibinfo {title} {Langevin simulations of the out-of-equilibrium dynamics of vortex glasses in high-temperature superconductors},}\ }\href {\doibase 10.1103/PhysRevB.75.024506} {\bibfield  {journal} {\bibinfo  {journal} {Phys. Rev. B}\ }\textbf {\bibinfo {volume} {75}},\ \bibinfo {pages} {024506} (\bibinfo {year} {2007})}\BibitemShut {NoStop}%
\bibitem [{\citenamefont {Luo}\ and\ \citenamefont {Hu}(2007)}]{PhysRevLett.98.267002}%
  \BibitemOpen
  \bibfield  {author} {\bibinfo {author} {\bibfnamefont {Meng-Bo}\ \bibnamefont {Luo}}\ and\ \bibinfo {author} {\bibfnamefont {Xiao}\ \bibnamefont {Hu}},\ }\bibfield  {title} {\enquote {\bibinfo {title} {Depinning and creep motion in glass states of flux lines},}\ }\href {\doibase 10.1103/PhysRevLett.98.267002} {\bibfield  {journal} {\bibinfo  {journal} {Phys. Rev. Lett.}\ }\textbf {\bibinfo {volume} {98}},\ \bibinfo {pages} {267002} (\bibinfo {year} {2007})}\BibitemShut {NoStop}%
\bibitem [{\citenamefont {Koshelev}\ and\ \citenamefont {Kolton}(2011)}]{PhysRevB.84.104528}%
  \BibitemOpen
  \bibfield  {author} {\bibinfo {author} {\bibfnamefont {A.~E.}\ \bibnamefont {Koshelev}}\ and\ \bibinfo {author} {\bibfnamefont {A.~B.}\ \bibnamefont {Kolton}},\ }\bibfield  {title} {\enquote {\bibinfo {title} {Theory and simulations on strong pinning of vortex lines by nanoparticles},}\ }\href {\doibase 10.1103/PhysRevB.84.104528} {\bibfield  {journal} {\bibinfo  {journal} {Phys. Rev. B}\ }\textbf {\bibinfo {volume} {84}},\ \bibinfo {pages} {104528} (\bibinfo {year} {2011})}\BibitemShut {NoStop}%
\bibitem [{\citenamefont {Dobramysl}\ \emph {et~al.}(2013)\citenamefont {Dobramysl}, \citenamefont {Assi}, \citenamefont {Pleimling},\ and\ \citenamefont {Täuber}}]{Dobramysl2013}%
  \BibitemOpen
  \bibfield  {author} {\bibinfo {author} {\bibfnamefont {Ulrich}\ \bibnamefont {Dobramysl}}, \bibinfo {author} {\bibfnamefont {Hiba}\ \bibnamefont {Assi}}, \bibinfo {author} {\bibfnamefont {Michel}\ \bibnamefont {Pleimling}}, \ and\ \bibinfo {author} {\bibfnamefont {Uwe~C.}\ \bibnamefont {Täuber}},\ }\bibfield  {title} {\enquote {\bibinfo {title} {Relaxation dynamics in type-ii superconductors with point-like and correlated disorder},}\ }\href {\doibase 10.1140/epjb/e2013-31101-x} {\bibfield  {journal} {\bibinfo  {journal} {The European Physical Journal B}\ }\textbf {\bibinfo {volume} {86}},\ \bibinfo {pages} {228} (\bibinfo {year} {2013})}\BibitemShut {NoStop}%
\bibitem [{\citenamefont {Assi}\ \emph {et~al.}(2015)\citenamefont {Assi}, \citenamefont {Chaturvedi}, \citenamefont {Dobramysl}, \citenamefont {Pleimling},\ and\ \citenamefont {T\"auber}}]{PhysRevE.92.052124}%
  \BibitemOpen
  \bibfield  {author} {\bibinfo {author} {\bibfnamefont {Hiba}\ \bibnamefont {Assi}}, \bibinfo {author} {\bibfnamefont {Harshwardhan}\ \bibnamefont {Chaturvedi}}, \bibinfo {author} {\bibfnamefont {Ulrich}\ \bibnamefont {Dobramysl}}, \bibinfo {author} {\bibfnamefont {Michel}\ \bibnamefont {Pleimling}}, \ and\ \bibinfo {author} {\bibfnamefont {Uwe~C.}\ \bibnamefont {T\"auber}},\ }\bibfield  {title} {\enquote {\bibinfo {title} {Relaxation dynamics of vortex lines in disordered type-ii superconductors following magnetic field and temperature quenches},}\ }\href {\doibase 10.1103/PhysRevE.92.052124} {\bibfield  {journal} {\bibinfo  {journal} {Phys. Rev. E}\ }\textbf {\bibinfo {volume} {92}},\ \bibinfo {pages} {052124} (\bibinfo {year} {2015})}\BibitemShut {NoStop}%
\bibitem [{\citenamefont {Assi}\ \emph {et~al.}(2016)\citenamefont {Assi}, \citenamefont {Chaturvedi}, \citenamefont {Dobramysl}, \citenamefont {Pleimling},\ and\ \citenamefont {T\"auber}}]{Assi01112016}%
  \BibitemOpen
  \bibfield  {author} {\bibinfo {author} {\bibfnamefont {Hiba}\ \bibnamefont {Assi}}, \bibinfo {author} {\bibfnamefont {Harshwardhan}\ \bibnamefont {Chaturvedi}}, \bibinfo {author} {\bibfnamefont {Ulrich}\ \bibnamefont {Dobramysl}}, \bibinfo {author} {\bibfnamefont {Michel}\ \bibnamefont {Pleimling}}, \ and\ \bibinfo {author} {\bibfnamefont {Uwe~C.}\ \bibnamefont {T\"auber}},\ }\bibfield  {title} {\enquote {\bibinfo {title} {Disordered vortex matter out of equilibrium: a langevin molecular dynamics study},}\ }\href {\doibase 10.1080/08927022.2015.1119826} {\bibfield  {journal} {\bibinfo  {journal} {Molecular Simulation}\ }\textbf {\bibinfo {volume} {42}},\ \bibinfo {pages} {1401--1409} (\bibinfo {year} {2016})}\BibitemShut {NoStop}%
\bibitem [{\citenamefont {Ambegaokar}\ \emph {et~al.}(1978)\citenamefont {Ambegaokar}, \citenamefont {Halperin}, \citenamefont {Nelson},\ and\ \citenamefont {Siggia}}]{PhysRevLett.40.783}%
  \BibitemOpen
  \bibfield  {author} {\bibinfo {author} {\bibfnamefont {Vinay}\ \bibnamefont {Ambegaokar}}, \bibinfo {author} {\bibfnamefont {B.~I.}\ \bibnamefont {Halperin}}, \bibinfo {author} {\bibfnamefont {David~R.}\ \bibnamefont {Nelson}}, \ and\ \bibinfo {author} {\bibfnamefont {Eric~D.}\ \bibnamefont {Siggia}},\ }\bibfield  {title} {\enquote {\bibinfo {title} {Dissipation in two-dimensional superfluids},}\ }\href {\doibase 10.1103/PhysRevLett.40.783} {\bibfield  {journal} {\bibinfo  {journal} {Phys. Rev. Lett.}\ }\textbf {\bibinfo {volume} {40}},\ \bibinfo {pages} {783--786} (\bibinfo {year} {1978})}\BibitemShut {NoStop}%
\bibitem [{\citenamefont {Huberman}\ \emph {et~al.}(1978)\citenamefont {Huberman}, \citenamefont {Myerson},\ and\ \citenamefont {Doniach}}]{PhysRevLett.40.780}%
  \BibitemOpen
  \bibfield  {author} {\bibinfo {author} {\bibfnamefont {B.~A.}\ \bibnamefont {Huberman}}, \bibinfo {author} {\bibfnamefont {R.~J.}\ \bibnamefont {Myerson}}, \ and\ \bibinfo {author} {\bibfnamefont {S.}~\bibnamefont {Doniach}},\ }\bibfield  {title} {\enquote {\bibinfo {title} {Dissipation near the critical point of a two-dimensional superfluid},}\ }\href {\doibase 10.1103/PhysRevLett.40.780} {\bibfield  {journal} {\bibinfo  {journal} {Phys. Rev. Lett.}\ }\textbf {\bibinfo {volume} {40}},\ \bibinfo {pages} {780--782} (\bibinfo {year} {1978})}\BibitemShut {NoStop}%
\bibitem [{\citenamefont {Myerson}(1978)}]{PhysRevB.18.3204}%
  \BibitemOpen
  \bibfield  {author} {\bibinfo {author} {\bibfnamefont {R.~J.}\ \bibnamefont {Myerson}},\ }\bibfield  {title} {\enquote {\bibinfo {title} {Quasiequilibrium statistical mechanics of two-dimensional superfluids and the two-dimensional coulomb gas},}\ }\href {\doibase 10.1103/PhysRevB.18.3204} {\bibfield  {journal} {\bibinfo  {journal} {Phys. Rev. B}\ }\textbf {\bibinfo {volume} {18}},\ \bibinfo {pages} {3204--3213} (\bibinfo {year} {1978})}\BibitemShut {NoStop}%
\bibitem [{\citenamefont {Reimann}\ \emph {et~al.}(2001)\citenamefont {Reimann}, \citenamefont {Van~den Broeck}, \citenamefont {Linke}, \citenamefont {H\"anggi}, \citenamefont {Rubi},\ and\ \citenamefont {P\'erez-Madrid}}]{PhysRevLett.87.010602}%
  \BibitemOpen
  \bibfield  {author} {\bibinfo {author} {\bibfnamefont {P.}~\bibnamefont {Reimann}}, \bibinfo {author} {\bibfnamefont {C.}~\bibnamefont {Van~den Broeck}}, \bibinfo {author} {\bibfnamefont {H.}~\bibnamefont {Linke}}, \bibinfo {author} {\bibfnamefont {P.}~\bibnamefont {H\"anggi}}, \bibinfo {author} {\bibfnamefont {J.~M.}\ \bibnamefont {Rubi}}, \ and\ \bibinfo {author} {\bibfnamefont {A.}~\bibnamefont {P\'erez-Madrid}},\ }\bibfield  {title} {\enquote {\bibinfo {title} {Giant acceleration of free diffusion by use of tilted periodic potentials},}\ }\href {\doibase 10.1103/PhysRevLett.87.010602} {\bibfield  {journal} {\bibinfo  {journal} {Phys. Rev. Lett.}\ }\textbf {\bibinfo {volume} {87}},\ \bibinfo {pages} {010602} (\bibinfo {year} {2001})}\BibitemShut {NoStop}%
\end{thebibliography}

%%%%%%%%% Merge with supplemental materials %%%%%%%%%%
\pagebreak
\newpage
\clearpage
\onecolumngrid
\begin{center}
\textbf{\large Supplemental Material}
\end{center}
%%%%%%%%%% Prefix a "S" to all equations, figures, tables and reset the counter %%%%%%%%%%
\setcounter{equation}{0}
\setcounter{figure}{0}
\setcounter{table}{0}
\setcounter{page}{1}
\makeatletter
\renewcommand{\theequation}{S\arabic{equation}}
\renewcommand{\thefigure}{S\arabic{figure}}
\renewcommand{\thetable}{S\arabic{table}}
\renewcommand{\bibnumfmt}[1]{[S#1]}
% \renewcommand{\citenumfont}[1]{S#1}
%%%%%%%%%% Prefix a "S" to all equations, figures, tables and reset the counter %%%%%%%%%%

\begin{center}
\textbf{I. Details of the 2D simulation for the BKT case}
\end{center}
In the BKT case, both vortices and antivortices are present. We denote the charge by $N_i$, where $N_i=+1$ for vortices and $N_i=-1$ for antivortices. The corresponding Lorentz force is given by $\mathbf{F}_L=N_i\,\mathbf{J}\times\mathbf{z}\Phi_0/c$. The viscous drag, random fluctuation, and pinning forces are independent of the vortex charge and therefore remain unchanged. Due to the renormalization effect, the prefactor $\frac{\Phi_0^2d}{8\pi^2\lambda^2}$ in the vortex–vortex interaction force should be replaced by $2\pi k_BTK_R$, where $K_R$ is the renormalized stiffness and $\pi K_R=2+s/2$. Additionally, the interaction force must be multiplied by $N_iN_j$ to distinguish between the repulsive interaction of like charges and the attractive interaction of opposite charges. Thus, the interaction force takes the form $\mathbf{F}_{vv}(\mathbf{r}_i-\mathbf{r}_j)=N_iN_j\frac{2\pi k_BTK_R}{\lambda}K_1(r_{ij}/\lambda)\frac{\mathbf{r}_i-\mathbf{r}_j}{r_{ij}}$.
We apply the same nondimensionalization procedure as in the critical field case, i.e., $\mathbf{r}_i \equiv \xi\, \tilde{\mathbf{r}}_i$, $t \equiv \frac{\lambda^3}{d\xi c^2 \rho_n}\,\tilde{t}$, $\mathbf{J} \equiv \frac{dc\Phi_0}{\lambda^3}\,\tilde{\mathbf{J}}$. With these substitutions, the Langevin equation Eq.(2) becomes
\begin{equation}
    \frac{\Delta \tilde{\mathbf{r}}_i}{\Delta \tilde{t}} = \frac{4\pi \lambda^2 k_BT\pi K_R}{d\Phi_0^2}\sum_{j\ne i}N_iN_j K_1\left(\frac{\tilde{r}_{ij}}{\kappa}\right)\frac{\tilde{\mathbf{r}}_i-\tilde{\mathbf{r}}_j}{\tilde{r}_{ij}} + 2\pi N_i\, \tilde{\mathbf{J}}\times \mathbf{z}
    + 2\sqrt{\frac{\pi k_BT\lambda^3}{d\xi\Phi_0^2\Delta\tilde{t}}}\,\boldsymbol{\sigma}\left(\tilde{t}\right) + \tilde{\mathbf{F}}_{pin}.\label{eq:Langevinnondim_RT}
\end{equation}

The density of vortices and antivortices depends on both current and temperature, and is given by Eq.(5a) divided by $2\pi\xi^2$. With this knowledge, Eq.\eqref{eq:Langevinnondim_RT} is ready to be put into simulations, using the same spatial configuration as in the critical field case. However, as the temperature decreases, the density becomes so low that fewer than one vortex and one antivortex are present in total within the finite sample, thus preventing meaningful numerical simulation. As a result, the lower-temperature portion of the numerical $R-T$ curves in Fig.~3 is estimated by substituting the velocity $|\tilde{v}|$, computed at the lowest numerically tractable temperature (where only one vortex and one antivortex are generated), into Eq.(4). These estimates are shown as dashed lines in the figure.

\begin{figure}[bh]
	\begin{center}
		\fig{3.6in}{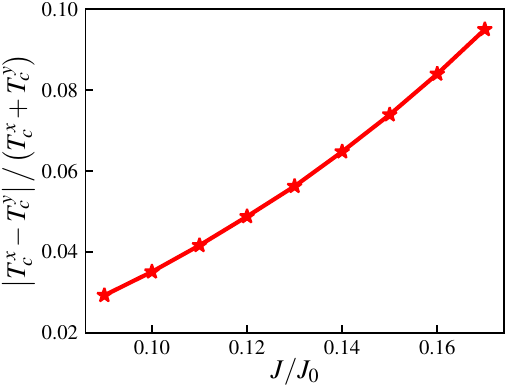}
		\caption{Analytical result for the dependence of $|T_c^x-T_c^y|/(T_c^x+T_c^y)$ on probing current. \label{fig:anatylical_current_BKT_anisotropy}}
	\end{center}
		%\vskip-1.5cm
\end{figure}

\begin{center}
    \textbf{II. Analytical treatment of the probing-current dependence of the BKT transition anisotropy}\label{sec:analytical_current_BKT_anisotropy}
\end{center}
In experiments, the critical temperature is typically identified as the temperature at which the resistance exceeds the minimum detectable value, $R_{\mathrm{min}}$. Since the resistance depends on the probing current, the observed values of $T_c^x$ and $T_c^y$ will vary with the applied current. To investigate how the normalized difference between the two critical temperatures depends on the probing current, we set Eq.(9) equal to $V_\mathrm{min}/(IR_n)$, where $V_{\mathrm{min}}$ is the minimum resolvable voltage, and solve it numerically. By substituting $U_0$ with $U_y$ and $U_x$, we obtain a series of values for $T_c^x$ and $T_c^y$ under different current magnitudes. The resulting relationship between $|T_c^x-T_c^y|/(T_c^x+T_c^y)$ and the probing current is shown in Fig.~\ref{fig:anatylical_current_BKT_anisotropy}. Obviously, the normalized difference increases with increasing current. To further examine the probing-current dependence beyond the simplified analytical treatment of the BKT case, and to include the mixed-state regime on the same footing, we additionally perform direct numerical simulations, as presented in Sec.~III.

\begin{figure}[t]
	\begin{center}
		\fig{4.6in}{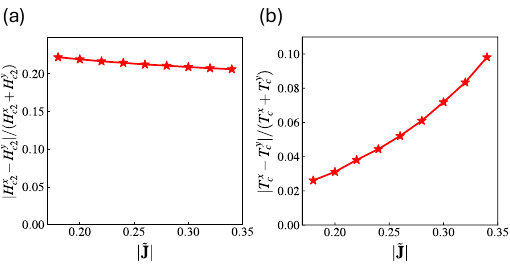}
		\caption{Numerical results for the probing-current dependence of the transition anisotropy. (a) Normalized mixed-state critical-field anisotropy $|H_{c2}^{x}-H_{c2}^{y}|/(H_{c2}^{x}+H_{c2}^{y})$ as a function of probing current. (b) Normalized BKT critical-temperature anisotropy $|T_{c}^{x}-T_{c}^{y}|/(T_{c}^{x}+T_{c}^{y})$ as a function of probing current.} \label{fig:varyJ}
	\end{center}
		%\vskip-1.5cm
\end{figure}

\begin{center}
    \textbf{III. Numerical simulations of the probing-current dependence of the transition anisotropy}\label{sec:numerical_current_dependence}
\end{center}

To complement the simplified analytical treatment of the zero-field BKT case in Sec.~II, we perform direct numerical simulations for both the finite-field mixed-state regime and the zero-field BKT regime. The corresponding results are summarized in Fig.~\ref{fig:varyJ}. In the mixed-state case, we characterize the transition anisotropy by the normalized difference $|H_{c2}^{x}-H_{c2}^{y}|/(H_{c2}^{x}+H_{c2}^{y})$. In the BKT case, we use the normalized difference $|T_{c}^{x}-T_{c}^{y}|/(T_{c}^{x}+T_{c}^{y})$.

For the finite-field mixed-state regime, the direct numerical results show that $|H_{c2}^{x}-H_{c2}^{y}|/(H_{c2}^{x}+H_{c2}^{y})$ decreases monotonically but only mildly with increasing probing current [Fig.~\ref{fig:varyJ}(a)]. This decreasing trend is consistent with the large-driving-force limit, where the pinning force becomes relatively less important and the directional distinction is expected to weaken. The weak magnitude of the decrease follows from Eq.~(4) of the main text: since the resistance is proportional to  $|\tilde{v}|/\tilde{J}$, the increase of vortex velocity with increasing driving force is partly compensated by the explicit division by $\tilde{J}$.

For the zero-field BKT regime, the direct numerical results show that $|T_{c}^{x}-T_{c}^{y}|/(T_{c}^{x}+T_{c}^{y})$ increases monotonically with increasing probing current [Fig.~\ref{fig:varyJ}(b)]. This behavior is qualitatively consistent with the simplified analytical result in Sec.~II. In this case, the dominant effect of increasing current is to generate more current-induced vortex-antivortex excitations, shifting the experimentally determined transition temperatures to lower values, where the pinning-induced suppression factor $1/[I_{0}(\beta U_{0})]^{2}$ in Eq.~(9) of the main text is more sensitive to the barrier height. This amplifies the difference between the two directional pinning barriers and increases the normalized anisotropy.
% \vspace{1.5em}

\begin{center}
    \textbf{IV. Neglect of transverse force terms in the vortex dynamics}\label{sec:transverse_force}
\end{center}

In the main text, we use a Langevin equation to describe the vortex dynamics. It retains the longitudinal viscous drag force, the Lorentz driving force, the vortex-vortex interaction, the pinning force, and thermal fluctuations, while neglecting transverse force terms. Here we clarify this approximation.

Following Kopnin~\cite{kopnin2001theory}, the general force balance on a moving vortex can be written as
\begin{equation}
  \mathbf{F}_{M}+\mathbf{F}_{L}^{(\mathrm{qp})}
  +\mathbf{F}_{I}+\mathbf{F}_{\mathrm{sf}}
  +\mathbf{F}_{\parallel}=0 ,
\end{equation}
where $\mathbf{F}_{M}$ is the Magnus force, $\mathbf{F}_{L}^{(\mathrm{qp})}$ is the Lorentz force from the quasiparticle current, $\mathbf{F}_{I}$ is the Iordanskii force, $\mathbf{F}_{\mathrm{sf}}$ is the spectral-flow force, and $\mathbf{F}_{\parallel}$ is the longitudinal force. The Magnus force has the form
\begin{equation}
  \mathbf{F}_{M}=\pi N_s \left[(\mathbf{v}_s-\mathbf{v}_L)\times\hat{\mathbf{z}}\right],
\end{equation}
where $\mathbf{v}_s$ is the superfluid velocity and $\mathbf{v}_L$ is the vortex velocity. The first part, $\pi N_s\mathbf{v}_s\times\hat{\mathbf{z}}$, together with
\begin{equation}
  \mathbf{F}_{L}^{(\mathrm{qp})}
  =\frac{\Phi_0}{c}\left[\mathbf{j}^{(\mathrm{qp})}\times\hat{\mathbf{z}}\right],
\end{equation}
gives the usual driving Lorentz force included in our formulation. The second part, $-\pi N_s\mathbf{v}_L\times\hat{\mathbf{z}}$, is perpendicular to the vortex velocity and belongs to the transverse-force sector, together with $\mathbf{F}_{I}$ and $\mathbf{F}_{\mathrm{sf}}$.

According to Kopnin~\cite{kopnin2001theory}, such transverse force terms become important only in the superclean limit $\omega_0\tau\gg 1$. The interface superconductor considered in the present work is not expected to be in this limit. Therefore, we neglect the transverse Magnus contribution as well as the other transverse-force terms. This approximation is consistent with our focus on the longitudinal vortex mobility controlled by the anisotropic pinning potential.

% \twocolumngrid

\end{document}